\documentclass{article}

\usepackage{microtype}
\usepackage{graphicx}
\usepackage{booktabs} 
\usepackage{makecell}
\usepackage{subcaption}

\usepackage{hyperref}

 \usepackage[accepted]{icml2024}

\usepackage{amsmath}
\usepackage{amssymb}
\usepackage{mathtools}
\usepackage{amsthm}
\usepackage{pifont}
\newcommand{\xmark}{\ding{55}}%

\usepackage[capitalize,noabbrev]{cleveref}

\theoremstyle{plain}
\newtheorem{theorem}{Theorem}[section]
\newtheorem{proposition}[theorem]{Proposition}
\newtheorem{lemma}[theorem]{Lemma}

\theoremstyle{definition}
\newtheorem{definition}[theorem]{Definition}

\theoremstyle{remark}
\newtheorem{remark}[theorem]{Remark}

\usepackage[textsize=tiny]{todonotes}

\icmltitlerunning{Learning Tube-Certified Control using Robust Contraction Metrics}

\begin{document}

\twocolumn[
\icmltitle{Learning Tube-Certified Control using Robust Contraction Metrics}



\icmlsetsymbol{equal}{*}

\begin{icmlauthorlist}
\icmlauthor{Vivek Sharma}{equal,yyy} 
\icmlauthor{Pan Zhao}{equal,comp} 
\icmlauthor{Naira Hovakimyan}{yyy}
\end{icmlauthorlist}

\icmlaffiliation{yyy}{Department of Mechanical Science and Engineering, University of Illinois Urbana-Champaign, Illinois, USA}
\icmlaffiliation{comp}{The University of Alabama, Alabama, USA}

\icmlcorrespondingauthor{Vivek Sharma}{viveks4@illinois.edu}

\icmlkeywords{Machine Learning, ICML}

\vskip 0.3in
]



\printAffiliationsAndNotice{\icmlEqualContribution} 

\begin{abstract}
Control design for general nonlinear robotic systems with guaranteed stability and/or safety in the presence of model uncertainties is a challenging problem. 
Recent efforts attempt to learn a controller and a certificate (e.g., a Lyapunov function or a contraction metric) jointly using neural networks (NNs), in which model uncertainties are generally ignored during the learning process. In this paper, for nonlinear systems subject to bounded disturbances, we present a framework for jointly learning a robust nonlinear controller and a  contraction metric using a novel disturbance rejection objective that certifies a tube bound using NNs for user-specified variables (e.g. control inputs). The learned controller aims to minimize the effect of disturbances on the actual trajectories of state and/or input variables from their nominal counterparts while providing certificate tubes around nominal trajectories that are guaranteed to contain actual trajectories in the presence of disturbances. Experimental results demonstrate that our framework can generate tighter (smaller) tubes and a controller that is computationally efficient to implement.


\end{abstract}

\section{Introduction}
\label{submission}

Learning-enabled control has demonstrated impressive performance in solving challenging control problems in robotics. However, such performance often comes with a lack of stability and/or safety guarantees, which prevents the learned controllers from being deployed to safety-critical systems. To resolve this issue, researchers have attempted to additionally learn a certificate alongside a controller using neural networks (NNs).  Such a certificate can be a Lyapunov function that certifies the stability of a fixed point \cite{Spencer2018LyapunovNN, 10.5555/3454287.3454579, 9400289, DaiH-RSS-21}, a contraction metric that certifies incremental stability, i.e., convergence to desired trajectories \cite{pmlr-v155-sun21b, 9115010, 9302618, 9691930}, or a barrier function that certifies set invariance \cite{9303785, pmlr-v164-dawson22a}, among others. Traditional methods for synthesizing these certificates often resort to the special structure of dynamics, e.g., strict feedback forms, or sum of squares (SOS) programming \cite{parrilo2000structured-sos,prajna2004nonlinear-sos} that is only applicable to polynomial dynamical systems of low and medium dimensions. In contrast, NN-based certificate synthesis is generally scalable to high-dimensional systems. Nevertheless, prevailing methods for generating certificates using NNs typically assume precise knowledge of the dynamics and are susceptible to performance degradation when confronted with model uncertainties. There is a need to synthesize {\it NN-based robust controllers} that minimize the effect of disturbances and certificates that establish the performance guarantee.
\begin{figure*}[h] 
    \centering
       \includegraphics[width=0.7\linewidth]{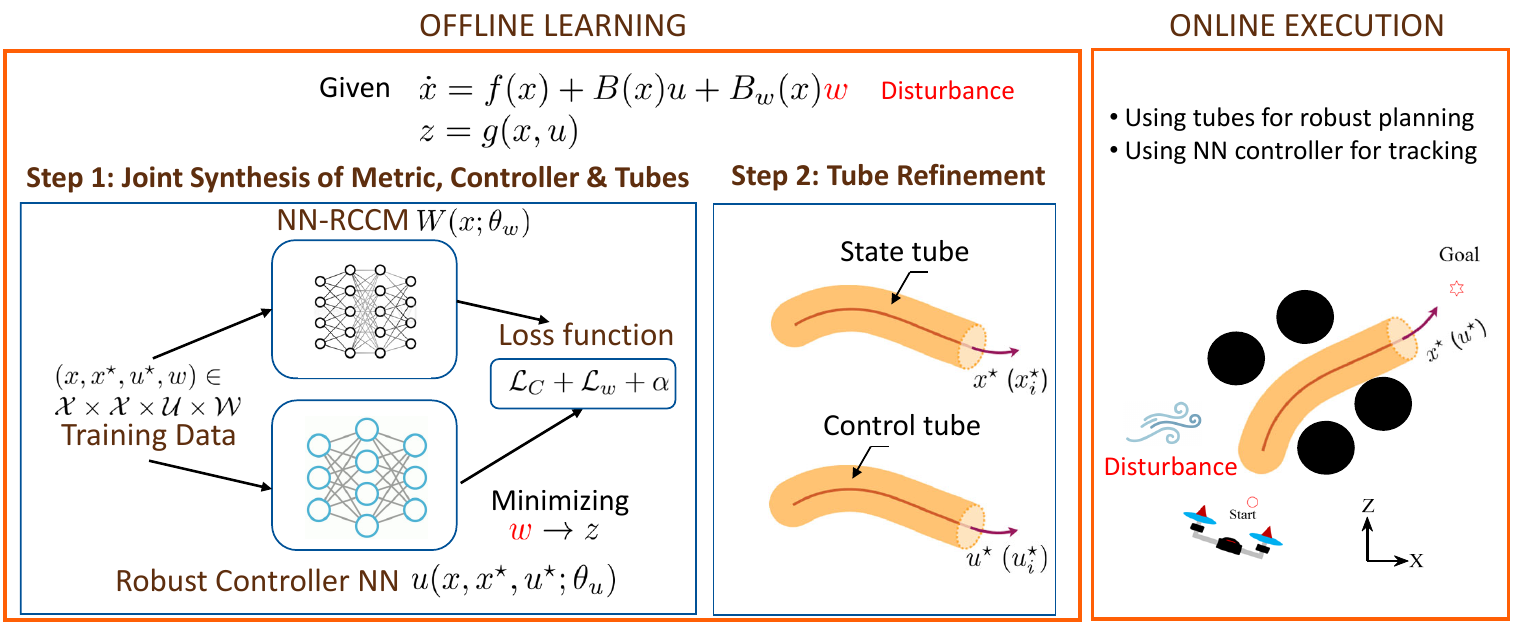}
  \caption{Our framework of jointly learning a robust controller and an RCCM for nonlinear disturbed systems, 
   while generating tubes to bound the difference between actual state/input trajectories ($x^\star/u^\star$) and their nominal counterparts. Offline phase involves (1) jointly learning the controller $u(x, x^*, u^*;\theta_u)$ and the metric $W(x;\theta_w)$ from data while minimizing tube size $\alpha$, and (2) refining tube size $\alpha$ for user specified variables (e.g., control inputs by setting $z=u$). During online execution, the tube can be leveraged by a motion planner to generate safe nominal trajectory $(x^*,u^*)$ that can be tracked by the controller $u(x, x^*, u^*;\theta_u)$. }
\label{fig:diag}
\vskip -0.1in
\end{figure*}

{\textbf{Related Work.}} \textbf{Contraction metrics}. Contraction theory~\cite{LOHMILLER1998683} provides a powerful tool for examining the incremental stability of nonlinear systems, i.e., the convergence between pairs of state trajectories towards each other, by analyzing the differential dynamics of the system. Recently, this theory has been extended for constructive control design through control contraction metrics (CCMs)~\cite{manchester2017control}. 
An attractive feature of the CCM-based nonlinear synthesis is that 
the search for a suitable CCM can be transformed into a convex optimization problem.  Contraction-based adaptive and robust control has also been investigated for nonlinear systems subject to unknown parameters \cite{lopez2020adaptive-ccm,gessow2023cdc}, state-dependent uncertainties \cite{lakshmanan2020safe,zhao2022robust-ccm-de}, and external disturbances \cite{zhao2022tube-ral,singh2019robust,manchester2018rccm}. In particular, \cite{zhao2022tube-ral} proposed a robust CCM to synthesize a robust nonlinear tracking controller that guarantees transient tracking performance via certificate tubes, which can be leveraged for safe planning under disturbances. However, to search for the (robust) CCMs, all the above approaches rely on SOS optimization or semidefinite programming (SDP), which does not scale well to high-dimensional systems~\cite{ahmadi2019dsos}.
 Additionally, to construct the control law, all the aforementioned approaches involve solving a nonlinear programming problem (to get the minimum energy path) at each time step, which is computationally heavy.


\textbf{Control certificate learning}. Certificate-based learning control differs from traditional learning control approaches, such as most reinforcement learning (RL) methods, which primarily focus on searching for a control policy. Instead, certificate-based methods simultaneously search for a control policy and a certificate that validates the properties of the closed-loop system such as stability and safety. NNs have proven to be powerful function approximators in learning control certificates, including Lyapunov functions~\cite{Spencer2018LyapunovNN, 10.5555/3454287.3454579, 9400289,DaiH-RSS-21,ganai2023learning}, control barrier functions~\cite{9303785, pmlr-v164-dawson22a}, and contraction metrics~\cite{pmlr-v155-sun21b, 9115010, 9302618, 9691930}, among others. Recent advancements have introduced a new class of RL methods that focus on jointly learning a control policy and certificate, as demonstrated in~\cite{Chang2021salc,Chow2018LyRL}. However, all these methods generally {\it do not explicitly account for disturbances} during the learning process. For a comprehensive overview of these methods, readers can refer to~\cite{annurev-control-042920-020211, dawson2022survey}.

{\textbf{Tube-based planning and predictive control}. Motion planning or predictive control for nonlinear uncertain systems with guaranteed safety is a challenging problem. 
Feedback motion planning (FMP) or tube model predictive control (MPC) aims to mitigate the effect of uncertainties through the use of an ancillary controller that tracks a nominal (or) desired trajectory. The ancillary controller typically provides tubes around nominal trajectories, which are guaranteed to contain actual trajectories despite uncertainties. Such tubes can be used to plan safety-guaranteed trajectories through constraint tightening. Different approaches have been proposed to synthesize the ancillary controller and its associated tube, such as local linearization \cite{althoff2014online-reachability,manchester2019robust-funnels}, sliding mode control \cite{lopez2018robust-sliding,lopez2019dynamic-tube-mpc}, LQR plus SOS verification \cite{tedrake2010lqrtree} and incremental Lyapunov function (iLF) \cite{kohler2020computationally-tube-mpc}. These approaches either need to re-compute the tube for each specific trajectory \cite{althoff2014online-reachability,manchester2019robust-funnels,tedrake2010lqrtree}, or apply to only a specific class of systems such as fully-actuated systems \cite{{lopez2018robust-sliding,lopez2019dynamic-tube-mpc}}, or necessitate the existence of a known iLF which is challenging to find for general nonlinear systems. In contrast, the recently proposed contraction metrics-based approaches \cite{zhao2022tube-ral,singh2019robust} are applicable to general nonlinear control-affine systems, and the metrics can be systematically synthesized using SDP. However, these approaches still suffer from the scalability issue of SDP and the high computational cost in implementing the controller, which motivates \cite{pmlr-v155-sun21b} and this work that aims to jointly learn the contraction metrics and controllers using NNs. NNs-based approximation methods have also been employed for (robust) MPC~\cite{paulson2020-rmpcapprox,nubert2020-mpctrack}. \cite{Fan-RSS-20} generally assumes the availability of a stabilizing policy and derives a probabilistic tube from data. In contrast, our approach does not rely on a pre-existing stabilizing policy assumption; instead, it learns such a policy while concurrently providing a tube to assess the policy's performance. Furthermore, our work can provide tubes for control inputs.
\cite{chou2021_cdc} outlines an approach to feedback-based motion planning for systems with unknown dynamics, leveraging deep control-affine approximations acquired from a dynamics dataset and optimizing a tracking-error bound while learning a controller. In contrast, our approach distinguishes itself by including a novel objective aimed at rejecting disturbances as part of the learning process. Other notable works considering robustness to bounded disturbances include Hamilton-Jacobi Bellman (HJB)-based methods discussed in ~\cite{choi2021robust, choi2023forward} which obtain Control Barrier Functions (CBFs) via HJB partial differential equations employing worst-case disturbance bounds.\cref{table:summary} provides a summary of the key characteristics of our approach and existing relevant approaches mentioned above.

\begin{table*}[t]
\caption{Summary of key characteristics of our approach compared to existing approaches.}
\label{table:summary}
\begin{center}
\begin{small}
\begin{sc}
\begin{tabular}{lcccc}
\toprule
\textbf{Controller} & \makecell{\textbf{Disturbance} \\ \textbf{Rejection}} & \makecell{\textbf{Minimizing Tube Size for}\\ \textbf{User-Specified Variables}} & \makecell{\textbf{Tube for}\\\textbf{ inputs}}& \makecell{\textbf{Computational} \\\textbf{Cost (Online)}}\\
\midrule
\textbf{NCM}~\cite{9115010} & \xmark & Difficult & \xmark  & Low\\
\textbf{NN-CCM}~\cite{pmlr-v155-sun21b} & \xmark & Difficult & \xmark  & Low\\
\textbf{RCCM}~\cite{zhao2022tube-ral} & \checkmark & Easy & \checkmark & High\\
\textbf{NN-RCCM (ours)} & \checkmark & Easy & \checkmark & Low\\
\bottomrule
\end{tabular}
\end{sc}
\end{small}
\end{center}
\vskip -0.2in
\end{table*}

\textbf{Statement of Contributions}. {For nonlinear control-affine systems subject to bounded disturbances, this work presents a novel framework to jointly learning a robust nonlinear controller with tubes,  and a contraction metric using NNs, as illustrated in \cref{fig:diag}. The learned controller aims to minimize the effect of disturbances on the deviations of actual state and/or input trajectories from their nominal counterparts and provide certificate tubes around nominal trajectories where actual ones are guaranteed to remain in the presence of disturbances. Compared to~\cite{pmlr-v155-sun21b}, our approach explicitly considers the disturbance rejection objective in learning the metric and controller and allows for optimizing the tube size for arbitrary combination of states and control inputs.  
Additionally, the controller yielded by our approach is computationally much cheaper to implement, compared to ~\cite{zhao2022tube-ral}, which necessitates solving a nonlinear programming problem to compute the control signal at each time step. 
To the best of our knowledge, this work represents the first attempt to use NNs to learn tube-certified nonlinear robust controllers with explicit disturbance rejection properties.} 

{\it Notations}. Let $\mathbb{R}^n$, $\mathbb{R}^{n \times n}$ and $\mathbb{R}^{+}$ denote an $n$-dimensional vector space, space of $n \times n$ matrices and set of positive real numbers respectively. We use the notation $A \succ 0 \ (A \prec 0)$ and $A \succeq 0 \ (A \preceq 0)$ to denote positive definite (negative definite) and positive semi-definite (negative semi-definite) symmetric matrices respectively. For a matrix valued function $M(x) : \mathbb{R}^{n} \rightarrow \mathbb{R}^{n \times n}$, its Lie derivative along a vector $v \in \mathbb{R}^n$ element wise, is computed as $\partial_v M(x):= \sum_i v^{i} \frac{\partial M}{\partial x^{i}}$. The notation $v^i$ is used to denote the \emph{i-th} element of a vector $v$. $\langle A \rangle $ is the shorthand notation for $A + A^T$. Also, $\|\cdot\|$ denotes the 2-norm of a vector or matrix. The notation $x\in \mathcal{L}_{\infty}$ indicates that $\|x(t)\|$ is bounded for all $t\geq0$. The $\mathcal{L}_{\infty}$ and truncated $\mathcal{L}_{\infty}$ norm of a function $x(t) : \mathbb{R}^{+} \rightarrow \mathbb{R}^n$ are defined as $\|x\|_{\mathcal{L}_\infty} \triangleq  \sup_{t\geq0} \|\mathcolor{blue}{x}\|$ and $\|x\|_{\mathcal{L}_\infty^{[0,T]}} \triangleq  \sup_{0\leq t\leq T} \|\mathcolor{blue}{x}\|$ respectively.

\section{Problem Statement and Preliminaries}
Consider a nonlinear control affine system of the form
\begin{equation}\label{eq:1} 
\begin{split}
    \dot{x}(t) &= f(x(t)) + B(x(t))u(t) + B_w(x(t))w(t) \\
    z(t) &= g(x(t), u(t)) ,
\end{split}
\end{equation}
where $x(t) \in  \mathcal{X} \subseteq \mathbb{R}^n$, $u(t) \in \mathcal{U} \subseteq \mathbb{R}^m$ and $w(t) \in \mathcal{W} \subseteq \mathbb{R}^l$ $\forall t \in \mathbb{R}^{+}$ are the vector of states, inputs and unknown disturbances, respectively. Here $\mathcal{X}$, $\mathcal{U}$, and $\mathcal{W}$ are compact sets representing state space, input space, and disturbance space respectively. The vector/matrix-valued functions $f(x)$, $g(x)$, $B(x)$, and $B_w(x)$ are known smooth functions of appropriate dimensions. The output variable $z(t) \in \mathbb{R}^p$ represents the variables whose deviation from the nominal value should be minimized. We use the notations $b_i$ and $b_{w,i}$ to represent \emph{ith} column of matrix $B$ and $B_w$ respectively.

For the system in \eqref{eq:1},  assume we have a nominal state trajectory $x^*(t)$ and input trajectory $u^*(t)$, satisfying the nominal dynamics
\begin{align}
    \dot{x}^{*}(t) &= f(x^{*}(t)) + B(x^{*}(t))u^{*}(t) + B_w(x^{*}(t))w^{*}(t) \nonumber \\
    z^{*}(t) &= g(x^{*}(t), u^{*}(t)), \label{eq:2} 
\end{align} 
where $w^*(t)$ is a vector of nominal disturbances (with $w^*(t) \equiv 0$ being a special case).

\noindent The goal of this paper is to learn a state-feedback controller for the system \eqref{eq:1} of the form
\begin{equation}
    u(t) = u^{*}(t) + k(x(t),x^*(t)) \label{eq:3}
\end{equation} 
that minimizes the gain from disturbance deviation ($w - w^{*}$) to output deviation ($z - z^{*}$) of the closed-loop system (obtained by plugging \eqref{eq:3} into \eqref{eq:1}) given by
\begin{equation}
\begin{split}
    \dot{x}(t) = &f(x(t)) + B(x(t))(u^{*}(t) + k(x(t),x^*(t))) \\
     & + B_w(x(t))w(t) \\
    z(t) = &  g(x(t), u^{*}(t) + k(x(t),x^*(t))) . \label{eq:4} 
\end{split}
\end{equation}

Specifically, such gain is quantified through the concept of {\em universal $\mathcal{L}_\infty$ gain} ~\cite{khalil2002nonlinear} as defined below.

\begin{definition}
\label{def:ugain}
The control system in \eqref{eq:4} achieves a universal $\mathcal{L}_\infty$ gain bound of $\alpha$, if for any target trajectory $x^*$, $w^*$ and $z^*$ satisfying \eqref{eq:4}, any initial condition $x(0)$ and any disturbance $w$ such that $w - w^*\in \mathcal{L}_{\infty}$,  for any $T\geq 0$, the condition
\begin{equation}
    \| z - z^*\|^2_{\mathcal{L}^{[0,T]}_{\infty}} \leq \alpha^2 \| w - w^*\|^2_{\mathcal{L}^{[0,T]}_{\infty}} + \beta(x(0),x^*(0)) \label{eq:5} 
\end{equation}
holds  for a function $\beta(x_1,x_2) \geq 0$ with $\beta(x,x) = 0.$
\end{definition}
\begin{remark}
The gain $\alpha$ in \cref{def:ugain}, in the similar spirit of tube size, is used to quantify the deviation of closed-loop trajectory $z(\cdot)$ from the nominal $z^*(\cdot)$ trajectory.
\end{remark}

\subsection{Robust Contraction Metrics}

Contraction theory~\cite{LOHMILLER1998683} analyzes the incremental stability of a system by studying the evolution of distance between two arbitrarily close trajectories. This theory applies Lyapunov conditions for studying the stability of the differential version of system \eqref{eq:1}. The differential dynamics of the system \eqref{eq:1} can be represented as:
\begin{equation}
\begin{split}
    \dot{\delta}_x &= A(x,u,w)\delta_x + B(x)\delta_u + B_w(x)\delta_w \\
    \delta_z &= C(x,u)\delta_x + D(x,u)\delta_u, \label{eq:6}
\end{split}
\end{equation} 
where $A(x,u,w) := \frac{\partial f}{\partial x}+\sum_{i=1}^{m}\frac{\partial b_i}{\partial x}u_i + \sum_{i=1}^{p}\frac{\partial b_{w,i}}{\partial x}w_i$, $C(x,u):= \frac{\partial g}{\partial x}$ and  $D(x,u):=\frac{\partial g}{\partial u}$. $\delta_x$, $\delta_u$ and $\delta_w$ denote the infinitesimal displacement between a pair of state, control, and disturbance trajectories respectively.

Likewise, the differential dynamics of closed-loop system \eqref{eq:4} can be obtained as:
\begin{equation}
\begin{split}
    \dot{\delta}_x &= \mathcal{A}\delta_x + \mathcal{B}\delta_w,\ \delta_z = \mathcal{C}\delta_x + \mathcal{D}\delta_w, \label{eq:7}
\end{split}
\end{equation} 
where ${\mathcal{A} \triangleq A + BK}$, $\mathcal{B} \triangleq B_w$, $\mathcal{C} \triangleq C + DK$ and $\mathcal{D} \triangleq 0$. Here $K(x,x^*) \triangleq \frac{\partial k}{\partial x}$ with $k$ representing the state-feedback part of the controller as defined in \eqref{eq:3}.

{Contraction theory introduces a method to quantify the virtual displacement ($\delta_x)$ between two arbitrarily close trajectories using a positive definite metric denoted as $M(x): \mathcal{X} \mapsto \mathbb{R}^{n \times n}$. This theory extends the principles of Lyapunov theory to study incremental stability by incorporating a differential analog of a Lyapunov function of the form $V(x, \delta_x) = \delta_x^T M(x) \delta_x$. By demonstrating that this function exponentially decreases, meaning $\dot{V}(x, \delta_x) \leq -2 \lambda {V}(x, \delta_x)$ for some positive constant $\lambda$, incremental exponential stability of the system can be established.}

{In~\cite{manchester2017control}, the authors present an important theorem for calculating a CCM using matrix inequalities. The theorem states that if a positive-definite metric $W(x)$ satisfies the following conditions for all $x$ and some $\lambda > 0$}
\begin{equation}
    B_{\perp}^T\left(\partial_{f}W(x) + \langle \frac{\partial f(x)}{\partial x} W(x) \rangle  + 2\lambda W(x)  \right) B_{\perp} \prec 0, 
    \label{eq:8}
\end{equation} 
\begin{equation}
        B_{\perp}^T\left(\partial_{b_j}W(x) - \langle \frac{\partial b_j(x)}{\partial x} W(x) \rangle \right) B_{\perp} = 0, \ j = 1,..,m ,\label{eq:9}
\end{equation} 
where $B_\perp$ is a matrix such that $B_\perp^T B = 0$ and $W(x) = M^{-1}(x)$ is the dual metric verifying $\underline{w}I \preceq W(x) \preceq \overline{w}I$,  with $\underline{w} = 1 \slash \overline{m}$ and $\overline{w} = 1 \slash \underline{m}$ , then there exists a tracking controller $k(x,x^*)$ such that closed loop trajectory $x(t)$ of the system \eqref{eq:4} exponentially converges to the nominal trajectory $x^*(t)$ of the system \eqref{eq:2}, with the rate $\lambda$.

The next lemma establishes {\em sufficient} conditions for a closed-loop system \eqref{eq:4} to admit a {\em guaranteed universal $\mathcal{L}_\infty$ gain}.

\begin{lemma}~\cite{zhao2022tube-ral}
\label{lem:conlemma}
The closed-loop system \eqref{eq:4} has a  universal $\mathcal{L}_\infty$ gain bound of $\alpha > 0$, if there exists a uniformly-bounded symmetric metric $\underline{m}I \preceq M(x) \preceq \overline{m}I$ with $0 < \underline{m} \leq \overline{m}$ and positive constants $\lambda$ and $\mu$, such that $\forall x,x^*, w$, we have:
\begin{align}
\begin{bmatrix}
 \langle M\mathcal{A} \rangle + \dot{M} + \lambda M  & M \mathcal{B} \\
                \mathcal{B}^T M & -\mu I_p \\
            \end{bmatrix} &\preceq 0 \label{eq:10} \\
 \begin{bmatrix}
 \lambda M & 0 \\
 0 & (\alpha - \mu)I_p \end{bmatrix} - \alpha^{-1}\begin{bmatrix}
                \mathcal{C}^T \\
                \mathcal{D}^T
            \end{bmatrix} \begin{bmatrix} \mathcal{C} & \mathcal{D} \end{bmatrix} &\succeq 0, \label{eq:11}
 \end{align} 
where  $\dot{M} = \sum_{i=1}^{n}\frac{\partial M}{\partial x_i} \dot{x}_i$ and $\dot{x}_i$ is given by \eqref{eq:4}. The matrices $\mathcal{A}$, $\mathcal{B}$, $\mathcal{C}$ and $\mathcal{D}$ are  defined in \eqref{eq:7}.
\end{lemma}
\begin{remark}
    We refer to the metric $M(x)$ in \cref{lem:conlemma} as a robust CCM (RCCM).
\end{remark}

\section{Learning Robust Controller, Contraction Metrics, and Tubes}

We use NNs to jointly learn a robust controller and an RCCM for the system \eqref{eq:1} while minimizing the universal $\mathcal{L}_{\infty}$ gain. Both the controller and metric are parameterized as neural networks and the parameters are optimized using loss functions inspired by contraction theory and \cref{lem:conlemma}. The training data for learning is sampled independently from the dataset ${\{(x_i, x^{*}_i,u^{*}_i,w_i) \in \mathcal{X} \times \mathcal{X} \times \mathcal{U} \times \mathcal{W} \}_{i=1}^N}$. The overall framework is shown in \cref{fig:diag}.

\subsection{Joint offline learning of the controller and RCCM}
\label{jlearn}

The controller $u(x,x^*,u^*;\theta_u)$ and the dual metric $W(x;\theta_w)$ are neural networks, parameterized by $\theta_u$ and $\theta_w$ respectively. We want to learn a controller and a metric that minimizes the $\mathcal{L}_{\infty}$ gain, $\alpha$. The gain quantifies the tube size in which the closed-loop system trajectories are bound to stay despite disturbances. Ideally, we would want the smallest tube size possible for the chosen state or input or a combination thereof with a given disturbance bound. We construct $u(x,x^*,u^*;\theta_u)$ to ensure  that if $x = x^* $ then $ u(x,x^*,u^*;\theta_u) = u^* \ \forall \ \theta_u$. Also $W(x,\theta_w)$ is a symmetric matrix by construction and $W(x,\theta_w) \succeq \underline{w}I, ~ \forall \ x$ and $\theta_w $. Here, $\underline{w}$ is a hyperparameter and is used to lower bound the smallest eigenvalue of the dual metric.


We denote the {\em LHS} of \eqref{eq:10} and \eqref{eq:11} from \cref{lem:conlemma} by  $C_1(x,x^*,u^*,w;\theta_u,\theta_w,\mu)$ and
 $C_2(x,x^*,u^*,w;\theta_u,\theta_w,\alpha,\mu)$ respectively. Let $\rho(S)$ denote the uniform distribution over the set $S$, where $S:= \mathcal{X} \times \mathcal{X} \times \mathcal{U} \times \mathcal{W}$. The {\em robust contraction risk} of the system is defined as follows:
\begin{equation}
\mathcal{L}_{C_1}(\theta_w,\theta_u,\mu) =\mathbb{E}_{(x,x^*,u^*,w)\sim \rho(S)} \  L_{PD} (-C_{1}(\cdot))
\label{eq:12} 
\end{equation} 
\begin{equation}
\hspace{-1mm}\mathcal{L}_{C_2}(\theta_w,\theta_u,\alpha,\mu)
= \mathbb{E}_{(x,x^*,u^*,w) \sim \rho(S)} \  L_{PD} (C_{2}(\cdot)), \label{eq:13} 
\end{equation}

\noindent where $L_{PD}(\cdot) \geq 0$ is a loss function used for penalizing the negative definiteness of its argument. $L_{PD}(A)= 0$ if and only if $A \succeq 0$. The optimal values of  $(\theta_w^*,\theta_u^*,\alpha^*,\mu^*)$ will ensure that the controller $u(x,x^*,u^*;\theta_u^*)$ and dual metric $W(x;\theta_w^*)$ satisfy \eqref{eq:11} and \eqref{eq:12} exactly, with  $\alpha^*$ being the optimal gain (or tube size).

To guide the optimization process, two auxiliary loss terms, inspired by the contraction theory \eqref{eq:8} and \eqref{eq:9} that define sufficient conditions for contraction, are used. Denoting the {\em LHS} of \eqref{eq:8} and \eqref{eq:9} by $C_3(x,\theta_w)$ and $\{C_4^{j}(x,\theta_w)\}_{j=1}^{m}$ respectively, the following risk functions are used:
\begin{align}  
\mathcal{L}_{w_1}(\theta_w) &= \mathbb{E}_{(x,x^*,u^*,w)\sim \rho(S)} \ L_{PD} (-C_{3}(\cdot)) \label{eq:14} \\
\mathcal{L}_{w_2}(\theta_w) &= \sum_{j=1}^{m} \mathbb{E}_{(x,x^*,u^*,w)\sim \rho(S)} \  \| C_4^{j} (\cdot)\|_{F},\label{eq:15} 
\end{align}
where $\|\cdot\|_F$ is the Frobenius norm.

Putting everything together, we have the following loss function to train the neural network using sampled data 
\begin{multline}
\hspace{-2mm}
\mathcal{L} (\theta_w,\theta_u,\alpha,\mu) = \frac{1}{N} \sum_{i=1}^{N}  L_{PD} (-C_{1}(\cdot)) \!+\! L_{PD} (C_{2}(\cdot))  + \\
     \! L_{PD} (-C_{3}(\cdot)) \!+ \!\sum_{j=1}^{m} \  \| C_4^{j} (\cdot)\|_{F} \! + \!\alpha, \label{eq:16}  
\end{multline}
where the training data $\{x_i,x^{*}_i,u^{*}_i,w_i\}_{i=1}^N$ is sampled independently from $\rho(S)$. {The arguments have been omitted for brevity. $L_{PD}$ is defined as follows: Given a matrix $X \in \mathbb{R}^{n \times n} $, $\xi$ number of points are randomly sampled from a unit norm ball i.e. $\{\eta_j \in \mathbb{R}^n \ | \  \|\eta_j\|_2 = 1\}_{j=1}^{\xi}$ and $L_{PD}$ is calculated as $L_{PD}(X) = \frac{1}{\xi} \sum_{j=1}^{\xi} \max\{0,-\eta_j^TX\eta_j\}$}.


\subsection{Refinement of state and control tubes}
\label{sec:refine}
When formulating the learning objective stated in \eqref{eq:16}, the primary focus is often on minimizing the universal $\mathcal{L}_{\infty}$ gain for the vector $z$ in \eqref{eq:1}. This vector $z$ contains weighted states and inputs, and the goal is to strike a balance between tracking performance and control efforts. Specifically, the vector $z$ can be represented as $z = [(Qx)^T, (Ru)^T]^T$, where $Q$ and $R$ are weighting matrices. Once the metric and controller have been learned, it is possible to obtain smaller tubes for various combinations of states, inputs, or both by appropriately selecting $g(x,u)$ in \eqref{eq:1} or matrices $C$ and $D$ in \eqref{eq:6}. This eliminates the need for retraining to optimize $(\theta_w,\theta_u)$ for the new variable $z$. The primary objective of the refinement process is to minimize $\alpha$ exclusively for the new $z$, utilizing the specified cost functions given by \eqref{eq:12} and \eqref{eq:13}, while maintaining the fixed values of the parameters $(\theta_w,\theta_u)$. The constraint of keeping the parameter $\theta_w$ fixed removes reliance on the costs outlined in \eqref{eq:14} and \eqref{eq:15}, which solely depend on $W(x;\theta_w)$. The optimization problem to refine $\alpha$ is solved offline, using the same learning framework, by detaching \eqref{eq:14} and \eqref{eq:15} from the computation graph and fixing $(\theta_w,\theta_u)$.


\subsection{Verification of matrix inequality conditions}
\label{verif}
Ensuring the 
stability and robustness of the closed-loop system can be achieved by finding a metric $M(x)$ and controller gain $K(x, x^*)$ that satisfy the matrix inequalities presented in \eqref{eq:10} and \eqref{eq:11} for all points in the uncountable set $S$. However, empirically verifying the satisfaction of these inequalities at every point within the uncountable set poses a significant challenge. Existing methods for neural network verification, such as mixed-integer linear programming~\cite{9304201} and satisfiability modulo theories (SMT)-based methods~\cite{10.5555/3454287.3454579}, have been proposed but are currently limited to small neural networks or require restrictive assumptions on neural network architectures, such as specific choices of activation functions. Other techniques for verifying NN controllers include statistical validation~\cite{nubert2020-mpctrack,karg2021-verifprob} and constrained learning~\cite{yankai2020-dnnmpc,yun2022-stcmpc}.  

Rigorous theoretical guarantees for the correctness of our learned metric through the satisfaction of matrix inequalities at every data point within the considered sets are indeed attainable. This can be achieved by computing the Lipschitz constants of the relevant variables and imposing stricter versions of the matrix inequalities to accommodate the gap between sampled points and an arbitrary point in the considered sets. However, the stricter conditions could be very conservative. It's also worth noting that recently proposed almost Lyapunov functions~\cite{shenyu2020almostlyapf} show that a system can still demonstrate stability even when the Lyapunov stability conditions are not satisfied at all points. 
Motivated by \cite{pmlr-v155-sun21b}, we present a method to provide rigorous theoretical guarantees of the satisfaction of the matrix inequalities \eqref{eq:10} and \eqref{eq:11} across a compact set through Lipschitz continuity. We provide a \cref{prop:1} showing that the largest eigenvalue of the left-hand side (LHS) of inequality \eqref{eq:10} and \eqref{eq:11} indeed has a Lipschitz constant if both the dynamics \eqref{eq:1} and the learned controller and metric have Lipschitz constants. These Lipschitz constants can then be utilized to discretize the state space into grids to further verify the validity of the matrix inequalities at each grid point.
The key idea for verifying that the matrix inequalities hold across all points in the compact set using the Lipschitz constants of their largest eigenvalues provided by  \cref{prop:1} is as follows. Consider a locally Lipschitz continuous function $f: \mathcal{X} \rightarrow \mathbb{R}$ with Lipschitz constant $L_f$,  i.e.,
\begin{equation}
    \|f(x^i)-f(x^j)\| \leq L_f \|{x^i-x^j}\|,\quad \forall x^i, x^j\in \mathcal X,
\end{equation}
 and a discretization of the domain $\mathcal{X}$ such that the distance between any grid point and its nearest neighbor is less than $\tau$. If $f(x^i) <  -L_f \tau$ holds for all grid point $x^i$, then $f(x) < 0$ holds for all $x \in \mathcal{X}$. Combining discrete samples and the Lipschitz constant from Proposition \ref{prop:1} can guarantee the strict satisfaction of matrix inequalities \eqref{eq:10} and \eqref{eq:11} on the uncountable set $\mathcal{S}$.
 
\begin{proposition}\label{prop:1}
Let $A$, $B$, $K$, and $M$ be functions of $x$, $x^{*}$, $u^{*}$ and $w$. If  $\dot{M}$, $M$, $A$, $B$, $B_w$, and $K$ have Lipschitz constants $L_{\dot{M}}$, $L_M$ , $L_A$, $L_B$, $L_{B_w}$ and $L_K$ respectively, and $2$-norms of the $M$, $A$, $B$, $C$, $D$, $B_w$, and $K$ are all bounded by $S_M , S_A, S_B , S_C, S_D, S_{B_w}$ and $S_K$ respectively and $\alpha > \mu$, then the largest eigenvalue function $\lambda_{max}\big(\text{LHS of Eq. }\eqref{eq:10}\big)$ and $\lambda_{max}\big(\text{LHS of Eq. }\eqref{eq:11}\big)$ for the two inequalities have Lipschitz constants $L_{\dot{M}} + 2 (S_M L_A + S_AL_M + S_M S_B L_K + S_B S_K L_M + S_M S_K L_B + \frac{\lambda}{2} L_M + \frac{1}{\mu}S_M L_M S_{B_w}^2 )$ and $\lambda L_M  +  \frac{2}{\alpha} S_C S_D L_K  + \frac{4}{\alpha} S_K S_D^2 L_K$, respectively.
\end{proposition}

\begin{figure*} [h]
\centering
\begin{tabular}{cccc}
\includegraphics[width=0.3\textwidth]{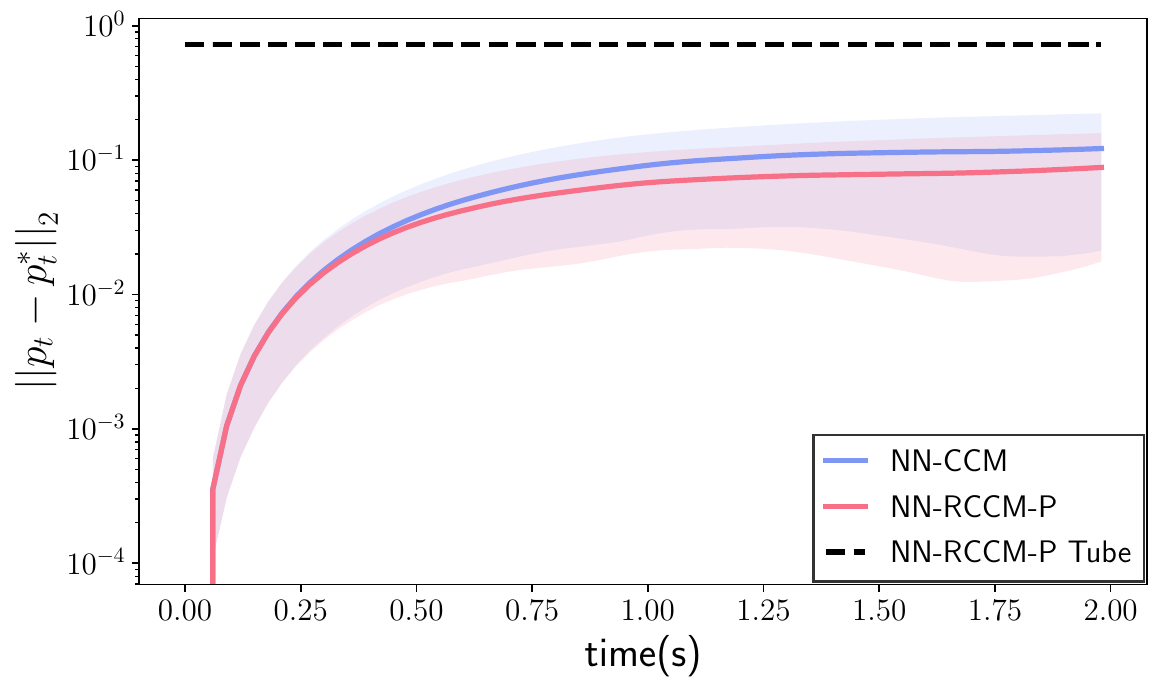} &
\includegraphics[width=0.3\textwidth]{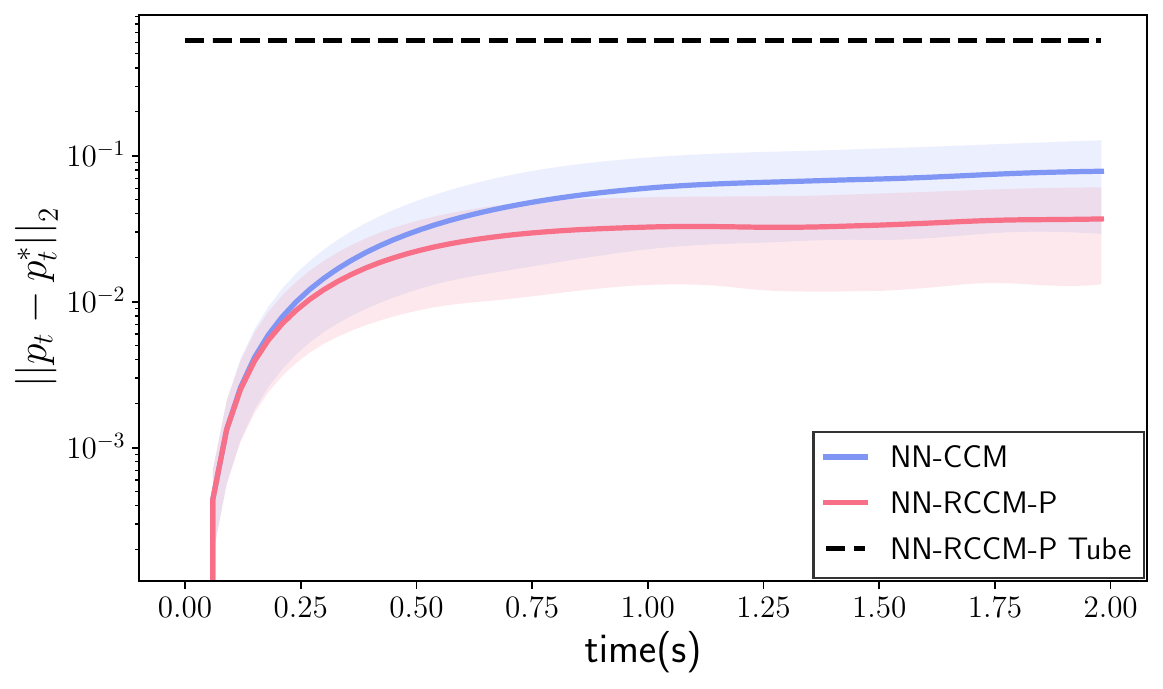} &
\includegraphics[width=0.3\textwidth]{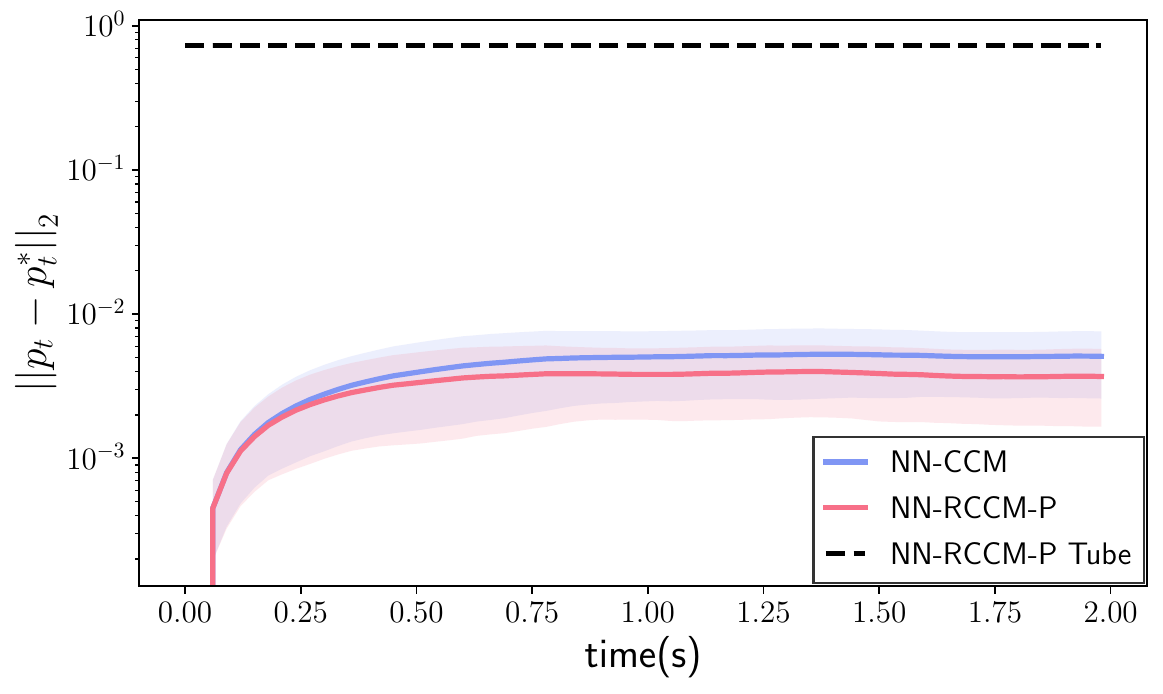} \\
(a) PVTOL & (b) QUADROTOR & (c) NL \\[6pt]
\end{tabular}
\begin{tabular}{cccc}
\includegraphics[width=0.3\textwidth]{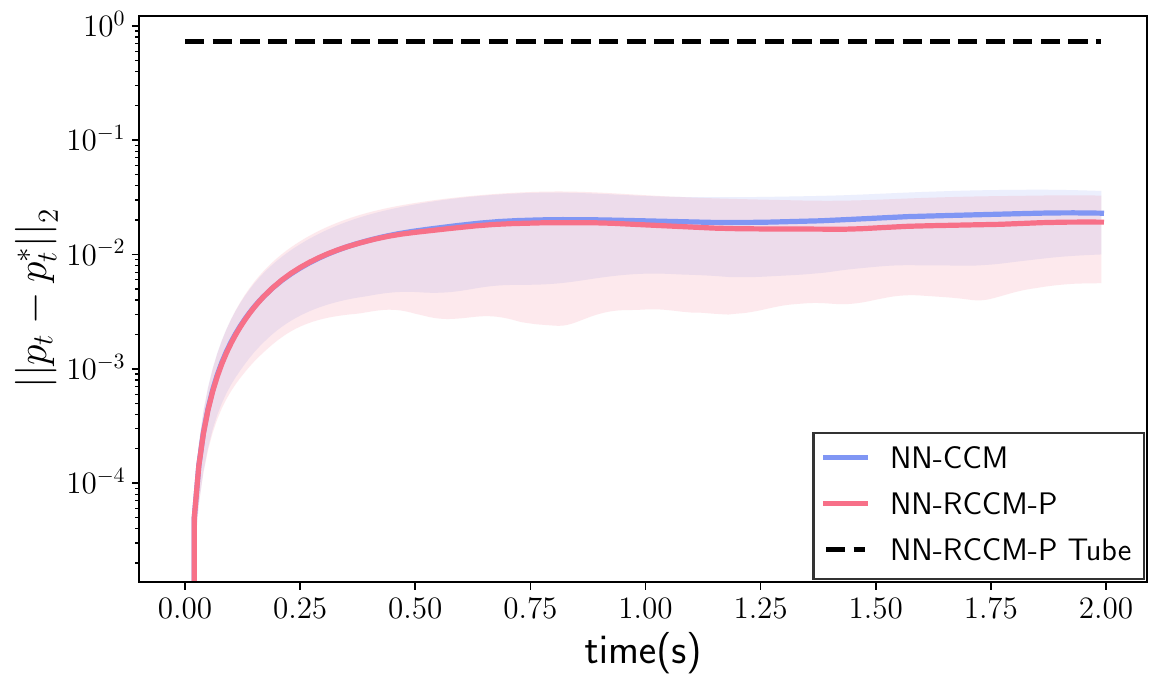} &
\includegraphics[width=0.3\textwidth]{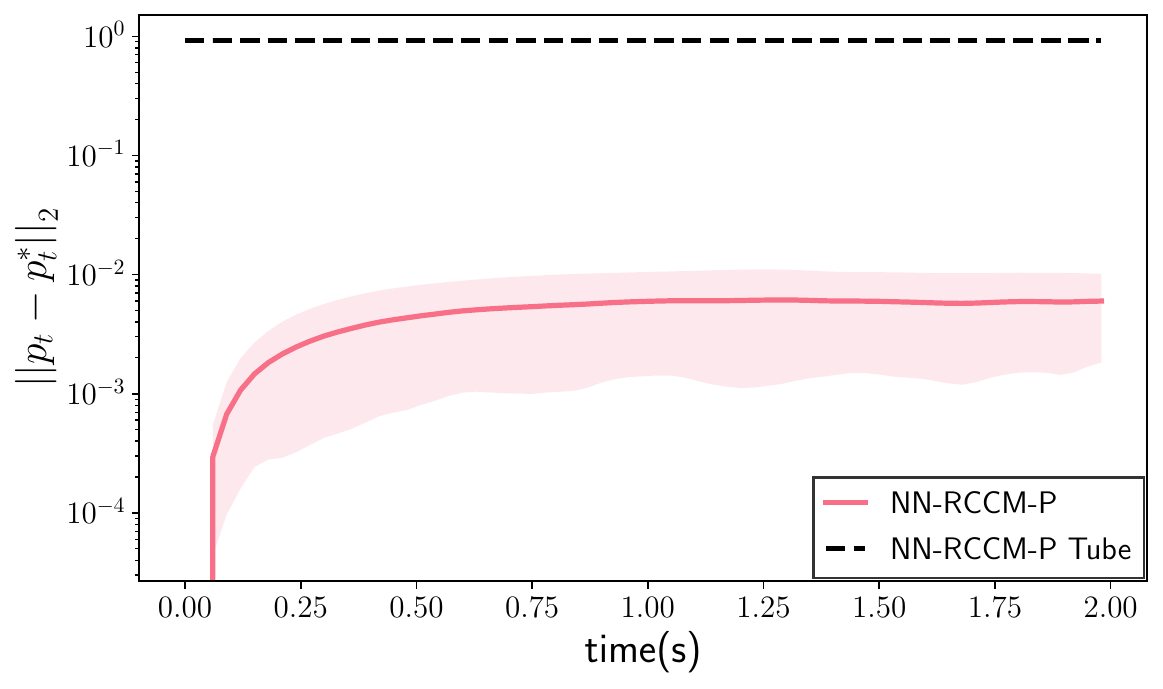} \\
(d) TLPRA  &  (e) 10-Link Rope  \\[6pt]
\end{tabular}
\caption{Tracking error comparisons for the four benchmark systems under {\bf NN-CCM} and {\bf NN-RCCM-P (ours)} in the presence of a disturbance with $\sigma = 1$. The $y$ axes are in log scale. The shaded regions are tracking errors between mean plus and minus one standard deviation over $100$ trajectories with the same initial conditions, i.e., $x(0) =x^*(0)$.}
\label{rccmp}
\end{figure*}

\section{Evaluation of Performance}
We test our framework on the four benchmark systems, namely, (1) a 4-dimensional planar vertical takeoff and landing vehicle \textbf{PVTOL} \cite{singh2019robust}, (2) a 10-dimensional \textbf{Quadrotor}~\cite{singh2019robust}}, (3) a 6-dimensional Neural Lander (\textbf{NL})~\cite{guyananl-2020} and (4) a 4-dimensional Two-Link Planar Robotic Arm (\textbf{TLPRA})~\cite{Yang-2012} (5) a 22-dimensional 10-link planar rope manipulation. The two benchmark systems, specifically NL and TLPRA, exhibit complex dynamics that cannot be effectively approximated by low-degree polynomials, making it challenging to apply SOS-based methods for control design. The 10-link rope manipulation benchmark~\cite{chou2021model-error} has been added to demonstrate the scalability of our framework to a system with a higher-dimensional state space and whose dynamics are unknown and can only be approximated from a set of expert demonstrations. \textbf{NN-CCM} and \textbf{RCCM-P} cannot handle such systems. 



To highlight the benefits of our framework in selectively optimizing tube sizes for specific states, we focus on learning a controller that minimizes tube sizes exclusively for the \emph{position states}. This approach is particularly suitable for motion planning tasks that prioritize obstacle avoidance. By reducing the tube sizes (and tracking errors) specifically for position states, collisions with obstacles can be effectively avoided. To achieve this, we introduce a controller referred to as \textbf{NN-RCCM-P}, where the function $g(x,u)$ is set to $p$, representing a vector that contains only the position states (e.g., $[p_x,p_z]$ for PVTOL).
For comparison, we designed a CCM-based neural controller in \cite{pmlr-v155-sun21b}, which we refer to as \textbf{NN-CCM}. Additionally, for PVTOL, we also designed a RCCM-based controller using the method described in~\cite{zhao2022tube-ral}, referred to as \textbf{RCCM-P}.

\subsection{Implementation Details}
\label{implement}
We utilize the neural network architecture and hyperparameter values used in~\cite{pmlr-v155-sun21b}. The dual metric $W(x;\theta_w)$ in our framework was modeled as $C(x;\theta_w)C^{T}(x;\theta_w) + \underline{w}I$, where $C(x;\theta_w)$ is a two-layer neural network with 128 neurons in the hidden layer, and $\underline{w}>0$ is a hyperparameter. The controller was constructed as $u(x,x^*,u^*,\theta_u) = u^* + \phi_2 \cdot \tanh(\phi_1 \cdot (x -x^*))$, where $\phi_1$ and $\phi_2$ are two two-layer neural networks with 128 neurons in the hidden layer with parameters denoted as $\theta_u = \{ \theta_{u1}, \theta_{u2} \}$ and $\tanh(\cdot)$ is the hyperbolic tangent function. The lowest eigenvalue of the dual metric is bounded by $\underline{w}$, and the optimization variables $\alpha$ and $\mu$ are initialized randomly and constrained to be positive. Both $\lambda$ and $\underline{w}$ are treated as hyperparameters.

For training, we randomly sample data points from the compact state space $\mathcal{X}$, control space $\mathcal{U}$, and disturbance space $\mathcal{W}$. Note that the training process is not limited by a specific structure for the nominal trajectory dataset, which means that our learned controller can track any nominal trajectory. Moreover, when simulating the nominal and closed-loop trajectories within the bounded time horizon $[0, T]$ for tracking error comparisons, we sampled the initial nominal state from the set $\mathcal{X}_0$ and the error between the initial nominal and actual states from a set $\mathcal{X}_{e0}$. We used the same compact set definitions {as described in~\cite{pmlr-v155-sun21b}} for defining the sets  $\mathcal{X}$, $\mathcal{U}$, $\mathcal{X}_0$ and $\mathcal{X}_{e0}$. The details are included in the \cref{sec:dyn}.
{Also, the disturbance vector $w$ is sampled from a compact set represented as} $\mathcal{W} := \{ w(t) \in \mathbb{R}^{p} | \|w\|_{\mathcal{L}_\infty} \leq \sigma \}$, where $\sigma$ is a constant denoting the bound of the disturbance. In simulations, we used a disturbance with a bound of $1$, i.e., $\sigma = 1$.


\begin{table}[t]
\caption{Total position tracking error. Mean $\pm$ standard deviation over $100$ trajectories with same initial conditions, i.e. $x(0) =x^*(0)$. RCCM-P method is not applicable for NL and TLPRA systems due to complex dynamics. }

\label{tpe}
\vskip 0.1in
\begin{center}
\begin{small}
\begin{sc}
\begin{tabular}{lcccc}
\toprule
\textbf{System} & \textbf{NN-CCM} & \makecell{\textbf{NN-RCCM-P}\\(Ours)} & \textbf{RCCM-P}\\
\midrule
PVTOL   & $0.074 \pm 0.047$ & $\mathbf{0.064} \pm 0.042$ & $0.1 \pm 0.05$ \\
Quad.& $0.051 \pm 0.026$ & $\mathbf{0.032} \pm 0.015$ & $0.16 \pm 0.03$\\
NL & $0.008 \pm 0.003$ & $\mathbf{0.007} \pm 0.002$ & NA\\
TLPRA & $0.034 \pm 0.017$ & $ \mathbf{0.030} \pm 0.013$ & NA\\
\bottomrule
\end{tabular}
\end{sc}
\end{small}
\end{center}
\vskip -0.2in
\end{table}



\subsubsection{Tracking error} \label{sec:tracker}
 The results are documented in \cref{rccmp} and \cref{tpe}. In our closed-loop simulations, we utilize a piecewise constant function to simulate the disturbance. For each interval of constant time, the length of the interval and the norm bound of the disturbance within that interval are uniformly sampled from the ranges $[0, 1]$ seconds and $[0.1, \sigma]$, respectively. In order to evaluate the tracking performance, we employ a quality metric referred to as the total tracking error. This metric is defined as follows: when presented with the tracking error curve $x_e(t)$ for $t \in [0, T]$, and given a specific $\sigma$ value as well as the initial condition $x(0) = x^*(0)$, we standardize the error curve by dividing it by the time duration $T$. The total tracking error is then represented by the area beneath this normalized error curve $x_e(t)/T$.



From \cref{rccmp} and \cref{tpe}, we can observe that the position tracking error for PVTOL, NL, and TLPRA is similar under both approaches. However, for Quadrotor, our \textbf{NN-RCCM-P}  yields a position tracking error that is approximately half of the error obtained with \textbf{NN-CCM}. Furthermore, it is important to highlight that the tracking error remains within the pre-computed bounds determined by the tube sizes. The tube bounds shown in ~\cref{rccmp} are larger than the tracking errors and this conservatism arises because our controller belongs to the category of robust controllers, which take into account worst-case scenarios.
\begin{figure}[ht] 
\vskip 0.1in
    \centering
    \includegraphics[width=0.7\linewidth]{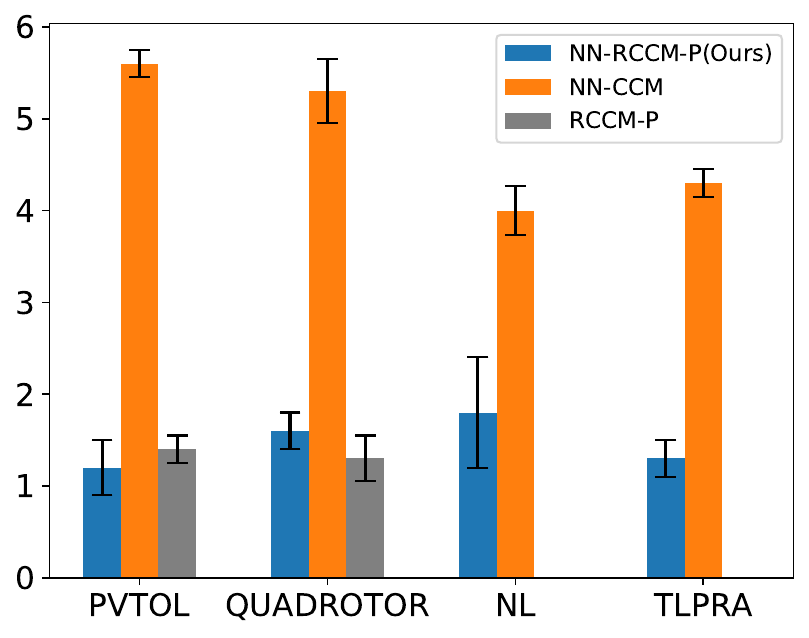}
  \caption{Tube size comparisons. Blue: tube size for position states under NN-RCCM-P. Orange: tube size for all states under NN-CCM. Gray : tube size for position states under RCCM-P. Note that the NN-CCM approach does not provide a way to compute tubes for individual states. RCCM-P method is not applicable for NL and TLPRA due to complex dynamics. The error bar ranges are one standard deviation of the tube sizes computed by varying the hyperparameter 
 $\lambda \in [0.5, 2.0]$ (contraction rate).}
\label{postube}
\vskip -0.1in
\end{figure}



\subsubsection{Tube sizes and execution times}
\cref{postube} presents a comparison of tube sizes for position states yielded by the two methods for the four benchmark systems.
In our method the tube size is determined by the $\mathcal{L}_{\infty}$-gain bound, $\alpha$. Conversely, for the \textbf{NN-CCM}-based method described in~\cite{pmlr-v155-sun21b}, the tube size is calculated using $\sqrt{\frac{\overline{w}}{\underline{w}}}\frac{\epsilon}{\lambda}$, where $\epsilon$ represents the upper bound on the disturbance term, satisfying $||B_w(x)w(t)||_2 \leq \epsilon$. To ensure a fair comparison, we tried to find the smallest possible tube size for the \textbf{NN-CCM} controller. For each value of $\lambda$, we fix $\underline{w}$ and then conduct a line search for the optimal $\overline{w}$ value which yields the smallest tube size. {Notably, our method yields much smaller tube sizes as compared to \textbf{NN-CCM} as seen in \cref{postube}.} 

\begin{figure}[ht] 
    \centering
    \includegraphics[width=0.7\linewidth]{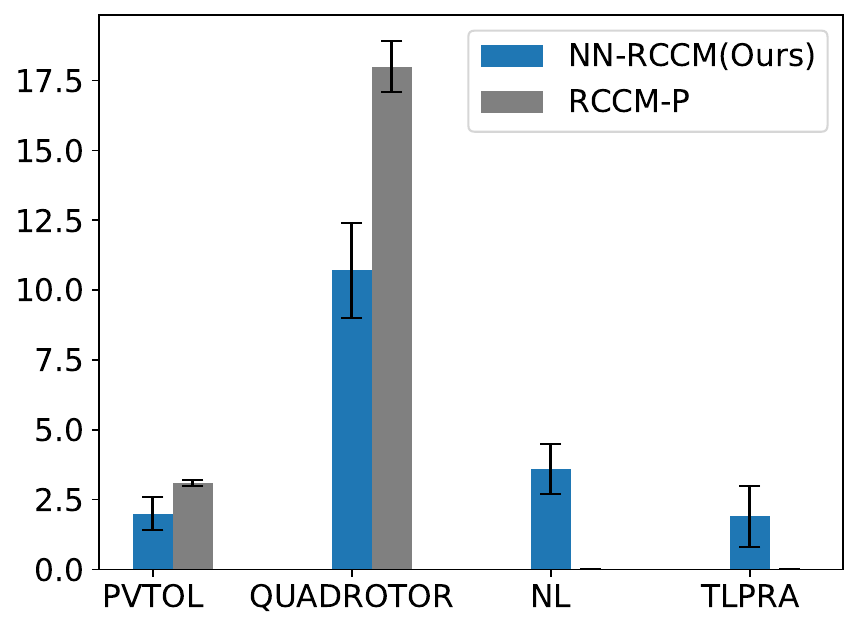}
  \caption{Tube size for control inputs. The error bars indicate one standard deviation of the tube sizes computed by varying hyperparameters  
 $\lambda \in [0.5, 2.0]$ (contraction rate). NN-CCM cannot provide such control tubes (as indicated in \cref{table:summary}). RCCM-P method is not applicable for NL and TLPRA due to complex dynamics.}
\label{ctube}
\end{figure}

In many motion planning applications, one must also know a tube for the control input to ensure control constraints are satisfied. Our framework facilitates the computation of control tube bounds after learning the metric and controller using the refinement approach elaborated in \cref{sec:refine}. Within \cref{ctube}, we show tube sizes for control
inputs yielded by our method for the four benchmark systems across a range of $\lambda$ values. These control tube bounds can be seamlessly incorporated for constraint tightening in motion planning applications. \textbf{NN-CCM}~\cite{pmlr-v155-sun21b} cannot provide such control tube bounds.



\begin{figure}[ht] 
    \centering
    \includegraphics[width=0.7\linewidth]{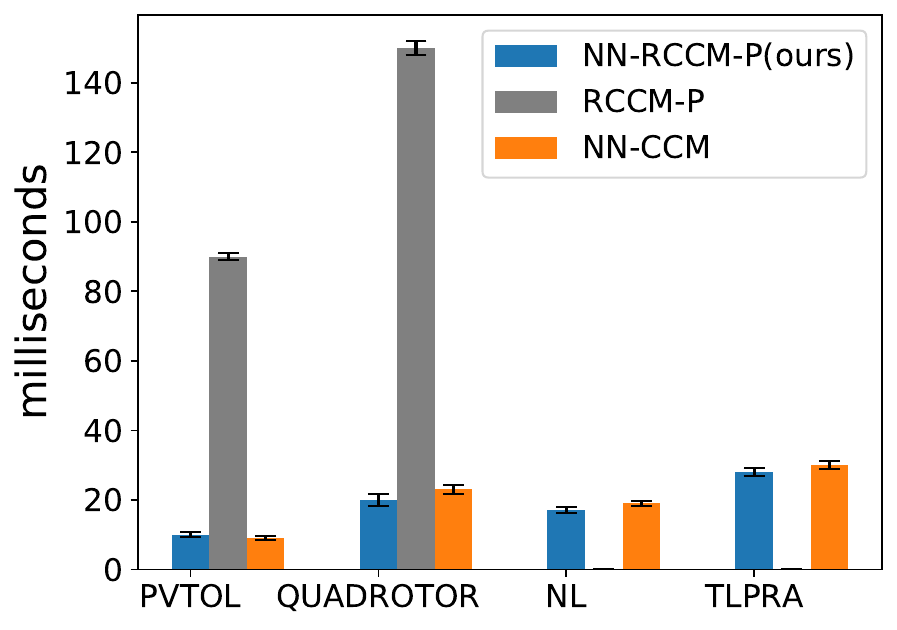}
  \caption{Execution time comparisons for controllers. The error bars indicate one standard deviation of the execution time computed across various nominal trajectory configurations. RCCM-P cannot find a stabilizing controller for NL and TLPRA.}
\label{exec}
\end{figure}

To demonstrate the advantage of our controller in reducing the computational overhead of implementation, we provide execution time comparisons (in milliseconds) of our framework with \textbf{RCCM-P} in \cref{exec}. The computational cost associated with our framework is $50$ times less than that associated with \textbf{RCCM-P}  as demonstrated in~\cref{exec}, as the latter involves solving a nonlinear programming (NLP) problem at each time to compute the geodesic for the control. Also, RCCM-P cannot find a stabilizing controller for the systems NL and TLPRA because their dynamics are too complex to be approximated by low-degree polynomials. In contrast, our NN-RCCM-P is capable of learning stabilizing controllers even for such complex systems, showcasing the scalability of our approach in designing stabilizing controllers for more complex systems. The reduced computational complexity provided by our method alleviates the burden of extensive computations, enabling agile and time-sensitive applications that were previously hindered by the computational demands of methods used in ~\cite{singh2019robust,zhao2022tube-ral}. It's worth mentioning that the execution time reported in \cref{exec} was obtained on a PC equipped with an Intel(R) i7-1065G7 CPU and 16 GB RAM. The time taken for offline training to learn \textbf{NN-CCM} and \textbf{NN-RCCM-P} on the benchmark systems is similar.

\begin{figure}[ht] 
 \vskip -0.1in
    \centering
    \includegraphics[width=0.7\linewidth]{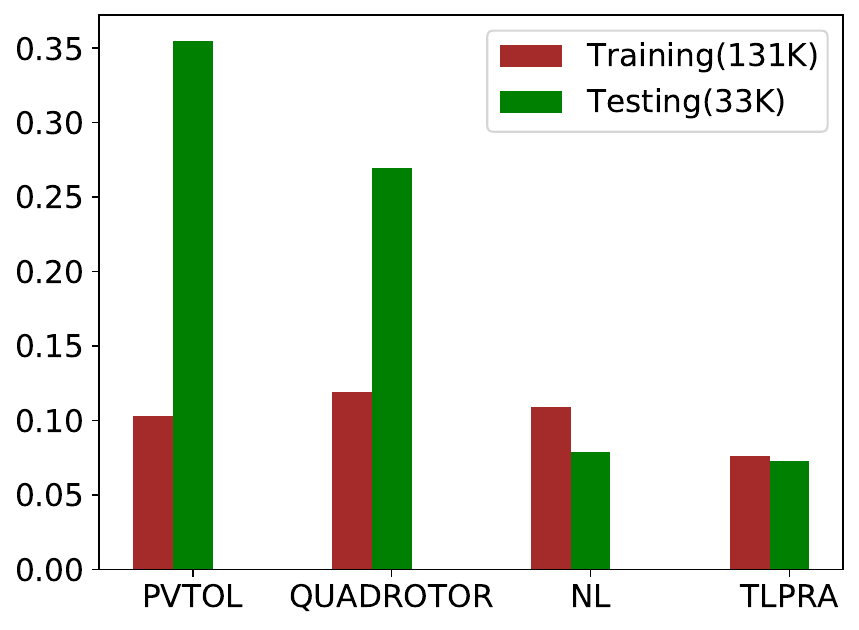}
  \caption{Fraction of total matrix inequality violations using our method during training and testing phases.}
\label{cert}
\vskip -0.2in
\end{figure}

 Our learning-based approach to stability certificates is established through the minimization of breaches in matrix inequalities at specific points. Our approach is suitable for generating stabilizing controllers for systems where ~\cite{zhao2022tube-ral} would fail. To ensure a fair comparison, we provide a statistical evaluation of the instances of violations in the four matrix inequalities employed for metric and controller learning within the four benchmark systems. This evaluation is carried out at the final iteration of training (upon the convergence of our $\mathcal{L}_{\infty}$-gain $\alpha$) and testing phase, as shown in~\cref{cert}. We achieved a level of tracking error performance comparable to \textbf{NN-CCM} based certificates but smaller tube sizes while experiencing almost the same frequency of certificate breaches.
\subsubsection{Tube-based motion planning}
\begin{figure}[ht]
 \vskip -0.1in
\begin{center}
\centerline{\includegraphics[width=0.35\textwidth]{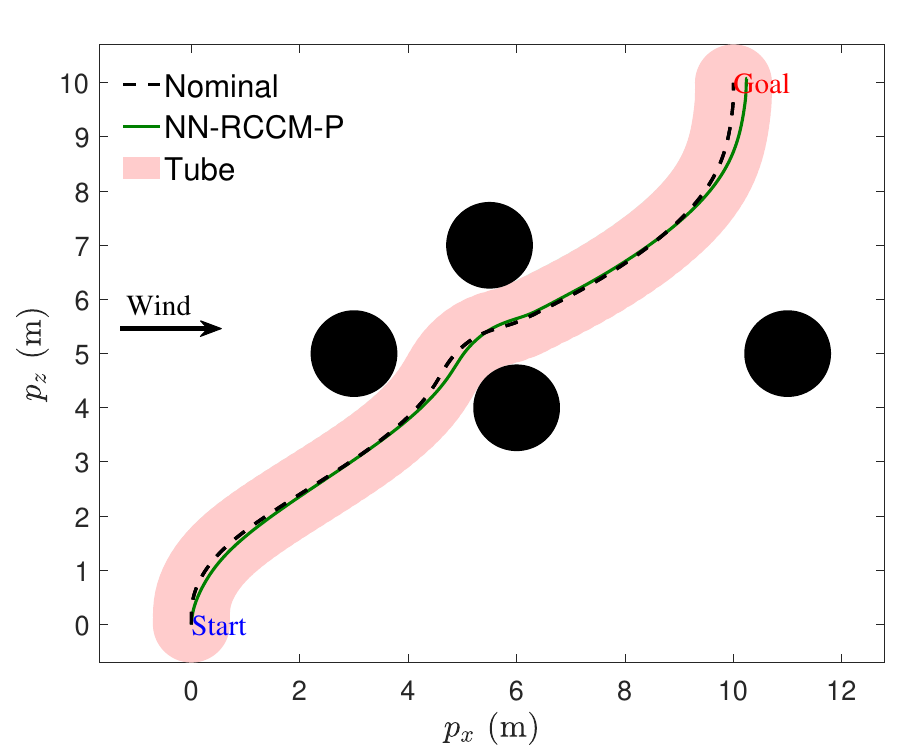}}
\caption{Trajectory planning and tracking for PVTOL under our controller \textbf{NN-RCCM-P}. Dotted line represents planned nominal trajectory, while shaded areas depict tubes for the position states. \textbf{NN-CCM} fails in identifying a feasible trajectory as the size of the associated tube is too large, making the goal ``unsafe''.}
\label{pvtolmp}
\end{center}
\vskip -0.2in
\end{figure}

We now demonstrate the integration of tubes for safe motion planning. We consider the task of maneuvering the PVTOL in the presence of obstacles toward the goal position. Employing our framework, we train a controller offline, optimizing tube size for both position states and control inputs. The online motion planner incorporates these tube sizes as constraints, generating a nominal trajectory that minimizes control effort and travel time while adhering to tube size constraints. Our controller then tracks this nominal trajectory, ensuring the actual system state remains within the tube despite bounded disturbances. Smaller tube sizes for position states enhance the efficiency of planning, particularly in confined spaces. Remarkably, our framework achieves a significantly tighter (smaller) tube for position states compared to NN-CCM. In contrast, NN-CCM yields large tube sizes for position states, rendering the computation of a nominal trajectory with tight position tube size constraints infeasible for safe maneuvering to the goal position, a task successfully undertaken by our framework, as illustrated in \cref{pvtolmp}.

\textbf{Limitations.} As noted earlier, our framework falls into the category of robust control that considers worst-case scenarios, which often lead to conservative performance (e.g. tubes shown in ~\cref{rccmp} are larger than the tracking errors). One possible way to mitigate the conservatism is to combine the proposed method with uncertainty compensation-based methods~\cite{lopez2020adaptive-ccm,lakshmanan2020safe-naira}, in which the matched uncertainties (that enter the system through the same channels as control inputs) can be estimated and compensated for, while the effect of unmatched uncertainties can be minimized using the proposed framework and we plan to explore it in future work. Additionally, the loss functions used to learn the NN metric and controller are soft relaxations of the constraints and this makes constraint satisfaction at every point in the state space challenging. Also, our approach of learning the metric and controller by penalizing the violations of soft constraints can be data inefficient and brittle in learning good controllers~\cite{richards2023learning}, although it provides attractive performance guarantees.





\section{Conclusions}
\label{sec:conclusion}
For nonlinear systems subject to disturbances, we have introduced a novel framework for joint learning of a robust contraction metric and a tube-certified controller with explicit {disturbance rejection capabilities} using neural networks. Our framework offers the flexibility to selectively optimize tube sizes for user-specified variables, resulting in a controller that generates tight tubes and is computationally cost-effective to implement. These tubes bound the difference between actual and nominal state/input trajectories and can be utilized by a motion planner to generate safety-guaranteed trajectories.

\section{Broader impact}
This paper presents work whose goal is to advance the field of Machine Learning. There are many potential societal consequences of our work, none of which we feel must be specifically highlighted here.

\nocite{langley00}

\bibliography{example_paper}

\begin{thebibliography}{51}
\providecommand{\natexlab}[1]{#1}
\providecommand{\url}[1]{\texttt{#1}}
\expandafter\ifx\csname urlstyle\endcsname\relax
  \providecommand{\doi}[1]{doi: #1}\else
  \providecommand{\doi}{doi: \begingroup \urlstyle{rm}\Url}\fi

\bibitem[Ahmadi \& Majumdar(2019)Ahmadi and Majumdar]{ahmadi2019dsos}
Ahmadi, A.~A. and Majumdar, A.
\newblock {DSOS and SDSOS} optimization: more tractable alternatives to sum of squares and semidefinite optimization.
\newblock \emph{SIAM Journal on Applied Algebra and Geometry}, 3\penalty0 (2):\penalty0 193--230, 2019.

\bibitem[Althoff \& Dolan(2014)Althoff and Dolan]{althoff2014online-reachability}
Althoff, M. and Dolan, J.~M.
\newblock Online verification of automated road vehicles using reachability analysis.
\newblock \emph{IEEE Transactions on Robotics}, 30\penalty0 (4):\penalty0 903--918, 2014.

\bibitem[Brunke et~al.(2022)Brunke, Greeff, Hall, Yuan, Zhou, Panerati, and Schoellig]{annurev-control-042920-020211}
Brunke, L., Greeff, M., Hall, A.~W., Yuan, Z., Zhou, S., Panerati, J., and Schoellig, A.~P.
\newblock Safe {L}earning in {R}obotics: {F}rom {L}earning-{B}ased {C}ontrol to {S}afe {R}einforcement {L}earning.
\newblock \emph{Annual Review of Control, Robotics, and Autonomous Systems}, 5\penalty0 (1):\penalty0 411--444, 2022.

\bibitem[Cao \& Gopaluni(2020)Cao and Gopaluni]{yankai2020-dnnmpc}
Cao, Y. and Gopaluni, R.~B.
\newblock Deep {N}eural {N}etwork {A}pproximation of {N}onlinear {M}odel {P}redictive {C}ontrol.
\newblock \emph{IFAC-PapersOnLine}, 53\penalty0 (2):\penalty0 11319--11324, 2020.
\newblock ISSN 2405-8963.
\newblock \doi{https://doi.org/10.1016/j.ifacol.2020.12.538}.
\newblock 21st IFAC World Congress.

\bibitem[Chang \& Gao(2021)Chang and Gao]{Chang2021salc}
Chang, Y.-C. and Gao, S.
\newblock Stabilizing {N}eural {C}ontrol {U}sing {S}elf-{L}earned {A}lmost {L}yapunov {C}ritics.
\newblock In \emph{2021 IEEE International Conference on Robotics and Automation (ICRA)}, pp.\  1803--1809, 2021.
\newblock \doi{10.1109/ICRA48506.2021.9560886}.

\bibitem[Chang et~al.(2019)Chang, Roohi, and Gao]{10.5555/3454287.3454579}
Chang, Y.-C., Roohi, N., and Gao, S.
\newblock \emph{Neural {L}yapunov {C}ontrol}.
\newblock Curran Associates Inc., Red Hook, NY, USA, 2019.

\bibitem[Chen et~al.(2021)Chen, Fazlyab, Morari, Pappas, and Preciado]{9400289}
Chen, S., Fazlyab, M., Morari, M., Pappas, G.~J., and Preciado, V.~M.
\newblock Learning {L}yapunov {F}unctions for {H}ybrid {S}ystems.
\newblock In \emph{2021 55th Annual Conference on Information Sciences and Systems (CISS)}, pp.\  1--1, 2021.
\newblock \doi{10.1109/CISS50987.2021.9400289}.

\bibitem[Choi et~al.(2021)Choi, Lee, Sreenath, Tomlin, and Herbert]{choi2021robust}
Choi, J.~J., Lee, D., Sreenath, K., Tomlin, C.~J., and Herbert, S.~L.
\newblock Robust control barrier--value functions for safety-critical control.
\newblock In \emph{2021 60th IEEE Conference on Decision and Control (CDC)}, pp.\  6814--6821. IEEE, 2021.

\bibitem[Choi et~al.(2023)Choi, Lee, Li, How, Sreenath, Herbert, and Tomlin]{choi2023forward}
Choi, J.~J., Lee, D., Li, B., How, J.~P., Sreenath, K., Herbert, S.~L., and Tomlin, C.~J.
\newblock A forward reachability perspective on robust control invariance and discount factors in reachability analysis.
\newblock \emph{arXiv preprint arXiv:2310.17180}, 2023.

\bibitem[Chou et~al.(2021{\natexlab{a}})Chou, Ozay, and Berenson]{chou2021_cdc}
Chou, G., Ozay, N., and Berenson, D.
\newblock Model {E}rror {P}ropagation via {L}earned {C}ontraction {M}etrics for {S}afe {F}eedback {M}otion {P}lanning of {U}nknown {S}ystems.
\newblock In \emph{2021 60th IEEE Conference on Decision and Control (CDC)}, pp.\  3576--3583, 2021{\natexlab{a}}.
\newblock \doi{10.1109/CDC45484.2021.9683354}.

\bibitem[Chou et~al.(2021{\natexlab{b}})Chou, Ozay, and Berenson]{chou2021model-error}
Chou, G., Ozay, N., and Berenson, D.
\newblock Model error propagation via learned contraction metrics for safe feedback motion planning of unknown systems.
\newblock In \emph{Proc. CDC}, pp.\  3576--3583, 2021{\natexlab{b}}.

\bibitem[Chow et~al.(2018)Chow, Nachum, Duenez-Guzman, and Ghavamzadeh]{Chow2018LyRL}
Chow, Y., Nachum, O., Duenez-Guzman, E., and Ghavamzadeh, M.
\newblock A {L}yapunov-based {A}pproach to {S}afe {R}einforcement {L}earning.
\newblock In Bengio, S., Wallach, H., Larochelle, H., Grauman, K., Cesa-Bianchi, N., and Garnett, R. (eds.), \emph{Advances in Neural Information Processing Systems}, volume~31. Curran Associates, Inc., 2018.

\bibitem[Dai et~al.(2020)Dai, Landry, Pavone, and Tedrake]{9304201}
Dai, H., Landry, B., Pavone, M., and Tedrake, R.
\newblock Counter-example guided synthesis of neural network {L}yapunov functions for piecewise linear systems.
\newblock In \emph{2020 59th IEEE Conference on Decision and Control (CDC)}, pp.\  1274--1281, 2020.
\newblock \doi{10.1109/CDC42340.2020.9304201}.

\bibitem[Dai et~al.(2021)Dai, Landry, Yang, Pavone, and Tedrake]{DaiH-RSS-21}
Dai, H., Landry, B., Yang, L., Pavone, M., and Tedrake, R.
\newblock {Lyapunov-stable neural-network control}.
\newblock In \emph{Proceedings of Robotics: Science and Systems}, Virtual, July 2021.
\newblock \doi{10.15607/RSS.2021.XVII.063}.

\bibitem[Dawson et~al.(2022)Dawson, Qin, Gao, and Fan]{pmlr-v164-dawson22a}
Dawson, C., Qin, Z., Gao, S., and Fan, C.
\newblock Safe {N}onlinear {C}ontrol {U}sing {R}obust {N}eural {L}yapunov-{B}arrier {F}unctions.
\newblock In Faust, A., Hsu, D., and Neumann, G. (eds.), \emph{Proceedings of the 5th Conference on Robot Learning}, volume 164 of \emph{Proceedings of Machine Learning Research}, pp.\  1724--1735. PMLR, 08--11 Nov 2022.

\bibitem[Dawson et~al.(2023)Dawson, Gao, and Fan]{dawson2022survey}
Dawson, C., Gao, S., and Fan, C.
\newblock Safe {C}ontrol with {L}earned {C}ertificates: {A} {S}urvey of {N}eural {L}yapunov, {B}arrier, and {C}ontraction {M}ethods for {R}obotics and {C}ontrol.
\newblock \emph{IEEE Transactions on Robotics}, pp.\  1--19, 2023.
\newblock \doi{10.1109/TRO.2022.3232542}.

\bibitem[Fan et~al.(2020)Fan, Agha, and Theodorou]{Fan-RSS-20}
Fan, D., Agha, A., and Theodorou, E.
\newblock {Deep {L}earning {T}ubes for {T}ube {MPC}}.
\newblock In \emph{Proceedings of Robotics: Science and Systems}, Corvalis, Oregon, USA, July 2020.
\newblock \doi{10.15607/RSS.2020.XVI.087}.

\bibitem[Ganai et~al.(2023)Ganai, Hirayama, Chang, and Gao]{ganai2023learning}
Ganai, M., Hirayama, C., Chang, Y.-C., and Gao, S.
\newblock Learning {S}tabilization {C}ontrol from {O}bservations by {L}earning {L}yapunov-like {P}roxy {M}odels.
\newblock In \emph{2023 IEEE International Conference on Robotics and Automation (ICRA)}, pp.\  2913--2920. IEEE, 2023.

\bibitem[Gessow \& Lopez(2023)Gessow and Lopez]{gessow2023cdc}
Gessow, S.~G. and Lopez, B.~T.
\newblock Adaptive {R}obust {C}ontrol {C}ontraction {M}etrics: {T}ransient {B}ounds in {A}daptive {C}ontrol with {U}nmatched {U}ncertainties.
\newblock In \emph{2023 62nd IEEE Conference on Decision and Control (CDC)}, pp.\  3541--3546, 2023.
\newblock \doi{10.1109/CDC49753.2023.10384247}.

\bibitem[Karg et~al.(2021)Karg, Alamo, and Lucia]{karg2021-verifprob}
Karg, B., Alamo, T., and Lucia, S.
\newblock Probabilistic performance validation of deep learning-based robust nmpc controllers.
\newblock \emph{International Journal of Robust and Nonlinear Control}, 31\penalty0 (18):\penalty0 8855--8876, 2021.
\newblock \doi{https://doi.org/10.1002/rnc.5696}.

\bibitem[Khalil(2002)]{khalil2002nonlinear}
Khalil, H.
\newblock \emph{Nonlinear Systems}.
\newblock Pearson Education. Prentice Hall, 2002.
\newblock ISBN 9780130673893.
\newblock URL \url{https://books.google.com/books?id=t_d1QgAACAAJ}.

\bibitem[K{\"o}hler et~al.(2020)K{\"o}hler, Soloperto, M{\"u}ller, and Allg{\"o}wer]{kohler2020computationally-tube-mpc}
K{\"o}hler, J., Soloperto, R., M{\"u}ller, M.~A., and Allg{\"o}wer, F.
\newblock A computationally efficient robust model predictive control framework for uncertain nonlinear systems.
\newblock \emph{IEEE Transactions on Automatic Control}, 66\penalty0 (2):\penalty0 794--801, 2020.

\bibitem[Lakshmanan et~al.(2020{\natexlab{a}})Lakshmanan, Gahlawat, and Hovakimyan]{lakshmanan2020safe}
Lakshmanan, A., Gahlawat, A., and Hovakimyan, N.
\newblock Safe feedback motion planning: {A} contraction theory and $\mathcal{L}_1$-adaptive control based approach.
\newblock In \emph{Proceedings of 59th IEEE CDC}, pp.\  1578--1583, 2020{\natexlab{a}}.

\bibitem[Lakshmanan et~al.(2020{\natexlab{b}})Lakshmanan, Gahlawat, and Hovakimyan]{lakshmanan2020safe-naira}
Lakshmanan, A., Gahlawat, A., and Hovakimyan, N.
\newblock Safe feedback motion planning: A contraction theory and $\mathcal{L}_1$-adaptive control based approach.
\newblock In \emph{Proc. CDC}, pp.\  1578--1583, 2020{\natexlab{b}}.

\bibitem[Langley(2000)]{langley00}
Langley, P.
\newblock Crafting papers on machine learning.
\newblock In Langley, P. (ed.), \emph{Proceedings of the 17th International Conference on Machine Learning (ICML 2000)}, pp.\  1207--1216, Stanford, CA, 2000. Morgan Kaufmann.

\bibitem[Li et~al.(2022)Li, Hua, and Cao]{yun2022-stcmpc}
Li, Y., Hua, K., and Cao, Y.
\newblock Using stochastic programming to train neural network approximation of nonlinear {MPC} laws.
\newblock \emph{Automatica}, 146:\penalty0 110665, 2022.
\newblock ISSN 0005-1098.
\newblock \doi{https://doi.org/10.1016/j.automatica.2022.110665}.

\bibitem[Liu et~al.(2020{\natexlab{a}})Liu, Shi, Chung, Anandkumar, and Yue]{guyananl-2020}
Liu, A., Shi, G., Chung, S.-J., Anandkumar, A., and Yue, Y.
\newblock Robust {R}egression for {S}afe {E}xploration in {C}ontrol.
\newblock In Bayen, A.~M., Jadbabaie, A., Pappas, G., Parrilo, P.~A., Recht, B., Tomlin, C., and Zeilinger, M. (eds.), \emph{Proceedings of the 2nd Conference on Learning for Dynamics and Control}, volume 120 of \emph{Proceedings of Machine Learning Research}, pp.\  608--619. PMLR, 10--11 Jun 2020{\natexlab{a}}.

\bibitem[Liu et~al.(2020{\natexlab{b}})Liu, Liberzon, and Zharnitsky]{shenyu2020almostlyapf}
Liu, S., Liberzon, D., and Zharnitsky, V.
\newblock Almost {L}yapunov functions for nonlinear systems.
\newblock \emph{Automatica}, 113:\penalty0 108758, 2020{\natexlab{b}}.
\newblock ISSN 0005-1098.
\newblock \doi{https://doi.org/10.1016/j.automatica.2019.108758}.

\bibitem[Lohmiller \& Slotine(1998)Lohmiller and Slotine]{LOHMILLER1998683}
Lohmiller, W. and Slotine, J.-J.~E.
\newblock On {C}ontraction {A}nalysis for {N}on-linear systems.
\newblock \emph{Automatica}, 34\penalty0 (6):\penalty0 683--696, 1998.
\newblock ISSN 0005-1098.
\newblock \doi{https://doi.org/10.1016/S0005-1098(98)00019-3}.

\bibitem[Lopez \& Slotine(2020)Lopez and Slotine]{lopez2020adaptive-ccm}
Lopez, B.~T. and Slotine, J.-J.~E.
\newblock Adaptive nonlinear control with contraction metrics.
\newblock \emph{IEEE Control Systems Letters}, 5\penalty0 (1):\penalty0 205--210, 2020.

\bibitem[Lopez et~al.(2018)Lopez, Slotine, and How]{lopez2018robust-sliding}
Lopez, B.~T., Slotine, J.-J., and How, J.~P.
\newblock Robust collision avoidance via sliding control.
\newblock In \emph{IEEE International Conference on Robotics and Automation (ICRA)}, pp.\  2962--2969, 2018.

\bibitem[Lopez et~al.(2019)Lopez, Slotine, and How]{lopez2019dynamic-tube-mpc}
Lopez, B.~T., Slotine, J.-J.~E., and How, J.~P.
\newblock Dynamic tube {MPC} for nonlinear systems.
\newblock In \emph{Proceedings of American Control Conference}, pp.\  1655--1662, 2019.

\bibitem[Manchester \& Slotine(2017)Manchester and Slotine]{manchester2017control}
Manchester, I.~R. and Slotine, J.-J.~E.
\newblock Control contraction metrics: {C}onvex and intrinsic criteria for nonlinear feedback design.
\newblock \emph{IEEE Trans. Autom. Control.}, 62\penalty0 (6):\penalty0 3046--3053, 2017.

\bibitem[Manchester \& Slotine(2018)Manchester and Slotine]{manchester2018rccm}
Manchester, I.~R. and Slotine, J.-J.~E.
\newblock Robust control contraction metrics: A convex approach to nonlinear state-feedback {$H_\infty$} control.
\newblock \emph{IEEE Control Syst. Lett.}, 2\penalty0 (3):\penalty0 333--338, 2018.

\bibitem[Manchester \& Kuindersma(2019)Manchester and Kuindersma]{manchester2019robust-funnels}
Manchester, Z. and Kuindersma, S.
\newblock Robust direct trajectory optimization using approximate invariant funnels.
\newblock \emph{Autonomous Robots}, 43\penalty0 (2):\penalty0 375--387, 2019.

\bibitem[Nubert et~al.(2020)Nubert, Köhler, Berenz, Allgöwer, and Trimpe]{nubert2020-mpctrack}
Nubert, J., Köhler, J., Berenz, V., Allgöwer, F., and Trimpe, S.
\newblock Safe and {F}ast {T}racking on a {R}obot {M}anipulator: {R}obust {M}pc and {N}eural {N}etwork {C}ontrol.
\newblock \emph{IEEE Robotics and Automation Letters}, 5\penalty0 (2):\penalty0 3050--3057, 2020.
\newblock \doi{10.1109/LRA.2020.2975727}.

\bibitem[Parrilo(2000)]{parrilo2000structured-sos}
Parrilo, P.~A.
\newblock \emph{Structured Semidefinite Programs and Semialgebraic Geometry Methods in Robustness and Optimization}.
\newblock PhD thesis, Massachusetts Institute of Technology, 2000.

\bibitem[Paulson \& Mesbah(2020)Paulson and Mesbah]{paulson2020-rmpcapprox}
Paulson, J.~A. and Mesbah, A.
\newblock Approximate {C}losed-{L}oop {R}obust {M}odel {P}redictive {C}ontrol {W}ith {G}uaranteed {S}tability and {C}onstraint {S}atisfaction.
\newblock \emph{IEEE Control Systems Letters}, 4\penalty0 (3):\penalty0 719--724, 2020.
\newblock \doi{10.1109/LCSYS.2020.2980479}.

\bibitem[Prajna et~al.(2004)Prajna, Papachristodoulou, and Wu]{prajna2004nonlinear-sos}
Prajna, S., Papachristodoulou, A., and Wu, F.
\newblock {Nonlinear control synthesis by sum of squares optimization: A Lyapunov-based approach}.
\newblock In \emph{5th Asian control conference}, volume~1, pp.\  157--165. IEEE, 2004.

\bibitem[Rezazadeh et~al.(2022)Rezazadeh, Kolarich, Kia, and Mehr]{9691930}
Rezazadeh, N., Kolarich, M., Kia, S.~S., and Mehr, N.
\newblock Learning {C}ontraction {P}olicies {F}rom {O}ffline {D}ata.
\newblock \emph{IEEE Robotics and Automation Letters}, 7\penalty0 (2):\penalty0 2905--2912, 2022.
\newblock \doi{10.1109/LRA.2022.3145100}.

\bibitem[Richards et~al.(2018)Richards, Berkenkamp, and Krause]{Spencer2018LyapunovNN}
Richards, S.~M., Berkenkamp, F., and Krause, A.
\newblock The {L}yapunov {N}eural {N}etwork: {A}daptive {S}tability {C}ertification for {S}afe {L}earning of {D}ynamical {S}ystems.
\newblock In \emph{Proceedings of The 2nd Conference on Robot Learning (CoRL)}, volume~87, pp.\  466--476. PMLR, 2018.

\bibitem[Richards et~al.(2023)Richards, Slotine, Azizan, and Pavone]{richards2023learning}
Richards, S.~M., Slotine, J.-J., Azizan, N., and Pavone, M.
\newblock Learning control-oriented dynamical structure from data.
\newblock In \emph{International Conference on Machine Learning}, pp.\  29051--29062. PMLR, 2023.

\bibitem[Robey et~al.(2020)Robey, Hu, Lindemann, Zhang, Dimarogonas, Tu, and Matni]{9303785}
Robey, A., Hu, H., Lindemann, L., Zhang, H., Dimarogonas, D.~V., Tu, S., and Matni, N.
\newblock Learning {C}ontrol {B}arrier {F}unctions from {E}xpert {D}emonstrations.
\newblock In \emph{2020 59th IEEE Conference on Decision and Control (CDC)}, pp.\  3717--3724, 2020.
\newblock \doi{10.1109/CDC42340.2020.9303785}.

\bibitem[Singh et~al.(2019)Singh, Landry, Majumdar, Slotine, and Pavone]{singh2019robust}
Singh, S., Landry, B., Majumdar, A., Slotine, J.-J., and Pavone, M.
\newblock Robust feedback motion planning via contraction theory.
\newblock \emph{The International Journal of Robotics Research, {\em under review}}, 2019.

\bibitem[Sun et~al.(2021)Sun, Jha, and Fan]{pmlr-v155-sun21b}
Sun, D., Jha, S., and Fan, C.
\newblock Learning {C}ertified {C}ontrol {U}sing {C}ontraction {M}etric.
\newblock In Kober, J., Ramos, F., and Tomlin, C. (eds.), \emph{Proceedings of the 2020 Conference on Robot Learning}, volume 155 of \emph{Proceedings of Machine Learning Research}, pp.\  1519--1539. PMLR, 16--18 Nov 2021.

\bibitem[Tedrake et~al.(2010)Tedrake, Manchester, Tobenkin, and Roberts]{tedrake2010lqrtree}
Tedrake, R., Manchester, I.~R., Tobenkin, M., and Roberts, J.~W.
\newblock {LQR-trees: Feedback motion planning via sums-of-squares verification}.
\newblock \emph{The International Journal of Robotics Research}, 29\penalty0 (8):\penalty0 1038--1052, 2010.

\bibitem[Tsukamoto \& Chung(2021)Tsukamoto and Chung]{9115010}
Tsukamoto, H. and Chung, S.-J.
\newblock Neural {C}ontraction {M}etrics for {R}obust {E}stimation and {C}ontrol: {A} {C}onvex {O}ptimization {A}pproach.
\newblock \emph{IEEE Control Systems Letters}, 5\penalty0 (1):\penalty0 211--216, 2021.
\newblock \doi{10.1109/LCSYS.2020.3001646}.

\bibitem[Tsukamoto et~al.(2021)Tsukamoto, Chung, and Slotine]{9302618}
Tsukamoto, H., Chung, S.-J., and Slotine, J.-J.~E.
\newblock Neural {S}tochastic {C}ontraction {M}etrics for {L}earning-{B}ased {C}ontrol and {E}stimation.
\newblock \emph{IEEE Control Systems Letters}, 5\penalty0 (5):\penalty0 1825--1830, 2021.
\newblock \doi{10.1109/LCSYS.2020.3046529}.

\bibitem[Yang \& Jagannathan(2012)Yang and Jagannathan]{Yang-2012}
Yang, Q. and Jagannathan, S.
\newblock Reinforcement {L}earning {C}ontroller {D}esign for {A}ffine {N}onlinear {D}iscrete-{T}ime {S}ystems using {O}nline {A}pproximators.
\newblock \emph{IEEE Transactions on Systems, Man, and Cybernetics, Part B (Cybernetics)}, 42\penalty0 (2):\penalty0 377--390, 2012.
\newblock \doi{10.1109/TSMCB.2011.2166384}.

\bibitem[Zhao et~al.(2022{\natexlab{a}})Zhao, Guo, and Hovakimyan]{zhao2022robust-ccm-de}
Zhao, P., Guo, Z., and Hovakimyan, N.
\newblock Robust nonlinear tracking control with exponential convergence using contraction metrics and disturbance estimation.
\newblock \emph{Sensors}, 22\penalty0 (13):\penalty0 4743, 2022{\natexlab{a}}.

\bibitem[Zhao et~al.(2022{\natexlab{b}})Zhao, Lakshmanan, Ackerman, Gahlawat, Pavone, and Hovakimyan]{zhao2022tube-ral}
Zhao, P., Lakshmanan, A., Ackerman, K., Gahlawat, A., Pavone, M., and Hovakimyan, N.
\newblock Tube-{C}ertified {T}rajectory {T}racking for {N}onlinear {S}ystems with {R}obust {C}ontrol {C}ontraction {M}etrics.
\newblock \emph{IEEE Robotics and Automation Letters}, 7\penalty0 (2):\penalty0 5528--5535, 2022{\natexlab{b}}.
\newblock \doi{10.1109/LRA.2022.3153712}.

\end{thebibliography}
\bibliographystyle{icml2024}

\newpage
\appendix
\onecolumn
\section{Appendix}
\subsection{Additional information on hyperparameters}

Consistent with the methodology outlined in~\cite{pmlr-v155-sun21b}, we use $\lambda = 0.5$ and $\underline{w} = 0.1$ as the hyperparameter values in our study for tracking error comparisons in \cref{rccmp} for the PVTOL, Quadrotor, Neural Lander and TLPRA systems. Our methodology involves learning parameters $(\theta_w,\theta_u)$ for the metric and controller correspondingly, through joint minimization of pointwise violations in the matrix inequalities and $\mathcal{L}_{\infty}$-gain denoted as $\alpha$. We trained the neural network for $15$ epochs using the Adam optimizer on a training dataset of size $N_{train} = 130,000$ uniformly sampled from the set $S$.

The important hyperparameters utilized in our learning framework include $\underline{w}$, learning rate, and the optimizer. In our exploration of the sensitivity of hyperparameters, one parameter that emerged as particularly influential is the contraction rate, denoted as $\lambda$. This parameter governs the exponential rate of convergence at which trajectories approach one another.Should $\lambda$ take an excessively small value, the convergence becomes slow, leading to poor transient response. Conversely, using an excessively large value will make it challenging to satisfy the matrix inequality condition in \eqref{eq:10}, resulting in the algorithm’s failure to acquire a stabilizing controller. A suitable range for $\lambda$ falls within $[0.5,2.0]$.

Another hyper parameter denoted as ${\underline{w}}$, is used which acts as a lower bound on the metric's eigenvalue in our algorithm. The algorithm was insensitive to variations in this parameter's value.

Regarding the range of optimizers we examined, the Adam optimizer, utilized with a learning rate of 0.001, successfully learned parameters that demonstrated the least violations of the matrix inequalities. We adopted the best set of additional hyperparameters, such as the number of layers and neurons, from~\cite{pmlr-v155-sun21b}.

\subsection{Ablation study for {the initialization of $\alpha$}}
\label{ablation}

\begin{figure}[ht]
\begin{center}
\centerline{\includegraphics[width=0.45\columnwidth]{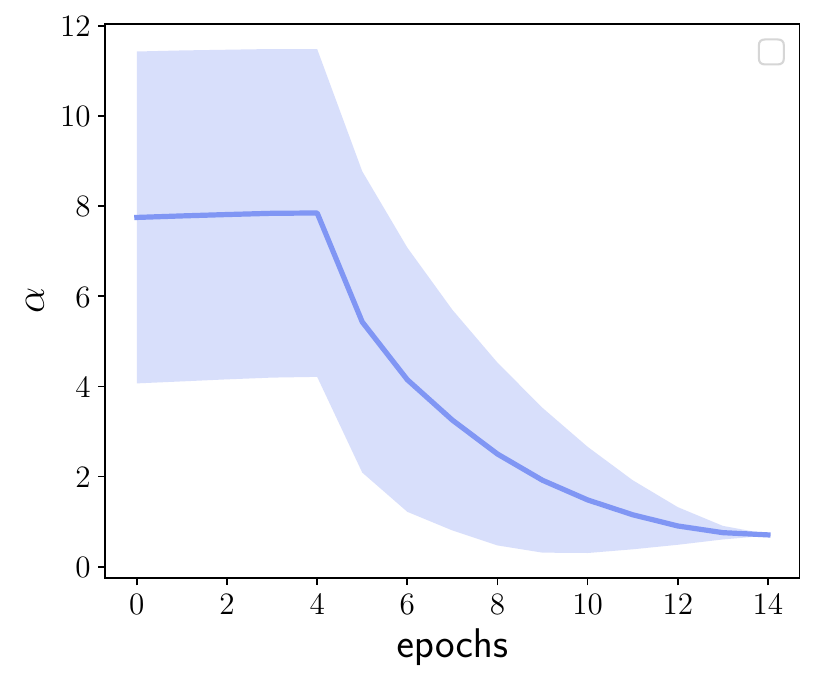}}
\caption{Convergence study of $\alpha$ with different initial values for optimizing the tube size for PVTOL system. The shaded region represents the $\alpha$ values between mean plus and minus one standard deviation computed for $20$ different initial values.}
\label{abalpha}
\end{center}
\end{figure}

Our loss function relies on the parameter $\alpha$. During the learning phase, we aim to minimize $\alpha$ while also penalizing violations of the robust contraction conditions to jointly learn the metric and controller. Consequently, the ultimate value that $\alpha$ converges to may be influenced by its initial value. To investigate this further, we conducted an ablation study on various initial values of $\alpha$ for the PVTOL system. The results, as depicted in~\cref{abalpha}, demonstrate that the convergent value of $\alpha$ remains unaffected by its initial value as the learning process progresses.

\newpage
\subsection{Dynamics}
\label{sec:dyn}
\subsubsection{PVTOL}
The PVTOL model, sourced from~\citep{singh2019robust}, represents a planar vertical take-off and landing aircraft. Its state vector, denoted as $x$, is defined as $x = [p_x, p_z, \phi, v_x, v_z, \dot{\phi}]^\top$, where $p = [p_x, p_z]^\top$ represents the position in the $x$ and $z$ directions respectively. Additionally, $v_x$ and $v_z$ correspond to the slip velocity (lateral) and the velocity along the thrust axis in the body frame of the vehicle, while $\phi$ denotes the angle between the $x$ direction of the body frame and the $x$ direction of the inertia frame. The input vector $u = [u_1, u_2]$ encompasses the thrust force generated by each of the two propellers. The dynamics governing the vehicle are expressed by
{\setlength{\arraycolsep}{1pt}
{\begin{equation*}
\small
   \dot{x} =
    \begin{bmatrix}
    v_x \cos(\phi)-v_z \sin(\phi)\\
    v_x\sin(\phi) + v_z \cos(\phi) \\
    \dot \phi \\
    v_z\dot \phi - g\sin(\phi) \\
    -v_x\dot \phi - g\cos(\phi) \\
    0  
    \end{bmatrix} +
    \begin{bmatrix}
    0 & 0 \\
    0 & 0 \\
    0 & 0 \\
    0 & 0 \\
    \frac{1}{m} & \frac{1}{m} \\
    \frac{l}{J} & -\frac{l}{J}
    \end{bmatrix}u + \begin{bmatrix}
     0 \\
     0 \\
     0\\
     \cos(\phi) \\
     -\sin(\phi)\\
      0
    \end{bmatrix}w,
\end{equation*}}}where $m$ and $J$ denote the mass and moment of inertia about the out-of-plane axis and $l$ is the distance between each of the propellers and the vehicle center, and $w\in\mathbb{R}$ denotes the disturbance in $x$ direction of the inertia frame. Following~\citep{pmlr-v155-sun21b}, the parameters were set as $m=0.486$ kg, $J=0.00383~\textup{Kg m}^2$, and $l=0.25$ m.

For the sets, we use $\mathcal{X} = \{x\in\mathbb{R}^6| [ -35, -2, -\pi/3 , -2, -1, -\pi/3 ] \leq x \leq [ 0, 2, \pi/3 , 2, 1, \pi/3 ] \}$, $\mathcal{U} = \{ u\in\mathbb{R}^2|[mg/2 - 1, mg/2 - 1] \leq u \leq [mg/2 + 1, mg/2 + 1]\}$, $\mathcal{X}_0 = \{x_0\in\mathbb{R}^6|[0, 0, -0.1, 0.5, 0, 0] \leq x_0 \leq [0, 0, 0.1, 1, 0, 0]\}$, and $\mathcal{X}_{e0} = \{x_{e0}\in\mathbb{R}^6| [-0.5, -0.5, -0.5, -0.5, -0.5, -0.5] \leq x_{e0} \leq [0.5, 0.5, 0.5, 0.5, 0.5, 0.5]\}$. 

\subsubsection{Quadrotor}
The 3D quadrotor model is taken from \cite{singh2019robust} and has the state-space representation given by $x=[p_x,p_y,p_z,v_x, v_y, v_z,f, \phi, \theta, \psi]^\top$, where the position $p=[p_x,p_y,p_z]^\top \in \mathbb{R}^3$ and corresponding velocities are expressed as $v=[v_x,v_y,v_z]^\top \in \mathbb{R}^3$.  Adopting the North-East-Down frame convention for the quadrotor body and the XYZ Euler-angle rotation sequence, the attitude (roll, pitch, yaw) is parameterized as
$(\phi, \theta,\psi)$ and $f >0$ is the total (normalized by mass) thrust generated by the four rotors. The control input is  $u = [\dot{f}, \dot \phi, \dot \theta,\dot \psi]^\top$. Given this parameterization, the dynamics of the quadrotor may be written as

{\setlength{\arraycolsep}{1pt}
{\begin{equation*}
\small
  \dot{x} =
    \begin{bmatrix}
    v_x \\
    v_y \\
    v_z \\
    -f \sin(\theta) \\
    f \sin(\theta)\cos(\phi) \\
    g - f \sin(\theta)\cos(\phi) \\
    0 \\
    0 \\
    0 \\
    0 \\
    \end{bmatrix} +
    \begin{bmatrix}
    0 & 0 & 0 & 0 \\
    0 & 0 & 0 & 0  \\
    0 & 0 & 0 & 0  \\
    0 & 0 & 0 & 0  \\
    0 & 0 & 0 & 0 \\
    0 & 0 & 0 & 0 \\
    1 & 0 & 0 & 0  \\
    0 & 1 & 0 & 0  \\
    0 & 0 & 1 & 0 \\
    0 & 0 & 0 & 1 \\
    \end{bmatrix}u + \begin{bmatrix}
     0  & 0 & 0\\
     0  & 0 & 0\\
     0  & 0 & 0\\
     1  & 0 & 0 \\
     0  & 1 & 0\\
     0  & 0 & 1\\
     0  & 0 & 0 \\
     0  & 0 & 0\\
     0  & 0 & 0 \\
     0  & 0 & 0\\
    \end{bmatrix}w,
\end{equation*}}}

where $g = 9.81$ and $w=[w_1,w_2,w_3]^\top\in\mathbb{R}^3$ denotes the disturbance.

Following~\citep{pmlr-v155-sun21b}, we use $\mathcal{X} = \{x\!\in\!\mathbb{R}^{10}|[-30, -30, -30, -1.5, -1.5, -1.5, 0.5g, -\pi/3, -\pi/3, -\pi/3] \\\leq x \leq [30, 30, 30, 1.5, 1.5, 1.5, 2g, \pi/3, \pi/3, \pi/3]\}$, $\mathcal{U} = \{u\in\mathbb{R}^4| [-1,-1,-1,-1] \leq u \leq   [1,1,\\1,1]\}$, $\mathcal{X}_0 = \{x_0\in\mathbb{R}^{10}|[-5, -5, -5, -1, -1, -1, g, 0, 0, 0] \leq x_0 \leq [5, 5, 5, 1, 1, 1, g, 0, 0, 0]\}$, and $\mathcal{X}_{e0} = \{x_{e0}\in\mathbb{R}^{10}|[-0.5, -0.5, -0.5, -0.5, -0.5, -0.5, -0.5, -0.5, -0.5, -0.5] \leq x_{e0} \leq [0.5, 0.5, 0.5, 0.5, 0.5, 0.5, 0.5, 0.5, 0.5, 0.5]\}$. 

\subsubsection{Neural Lander}
The Neural Lander model is taken from ~\cite{guyananl-2020} and has the state-space representation given by $x=[p_x,p_y,p_z,v_x, v_y, v_z]^\top$, where the position is given by $p=[p_x,p_y,p_z]^\top \in \mathbb{R}^3$ and the corresponding velocities are given as as $v=[v_x,v_y,v_z]^\top \in \mathbb{R}^3$.  The control input $u = [a_x,a_y,a_z]^\top$ corresponds to acceleration in three directions.
This model uses a neural-network $Fa_i$ to model ground effects using data ~\cite{guyananl-2020} and because a neural network cannot be approximated by low-degree polynomials, methods relying on low-degree polynomial approximation of dynamics fail to design a stabilizing controller. The system dynamics are defined as


{\setlength{\arraycolsep}{1pt}
{\begin{equation*}
\small
  \dot{x} =
    \begin{bmatrix}
    v_x \\
    v_y \\
    v_z \\
    Fa_1/m \\
    Fa_2/m \\
    Fa_3/m - g\\
    \end{bmatrix} +
    \begin{bmatrix}
    0 & 0 & 0  \\
    0 & 0 & 0   \\
    0 & 0 & 0  \\
    1 & 0 & 0   \\
    0 & 1 & 0  \\
    0 & 0 & 1 \\
    \end{bmatrix}u + \begin{bmatrix}
     0  & 0 & 0\\
     0  & 0 & 0\\
     0  & 0 & 0\\
     1  & 0 & 0 \\
     0  & 1 & 0\\
     0  & 0 & 1\\
    \end{bmatrix}w,
\end{equation*}}}

where $m = 1.47$, $g = 9.81$ and $Fa_i = Fa_i(z, v_x, v_y, v_z )$ for $i = 1, 2, 3$ are neural networks and $w=[w_1,w_2,w_3]^\top\in\mathbb{R}^3$ denotes the disturbance.

Following~\citep{pmlr-v155-sun21b}, we use $\mathcal{X} = \{x\!\in\!\mathbb{R}^{6}|[-5, -5, 0, -1, -1, -1] \\\leq x \leq [5, 5, 2, 1, 1, 1]\}$, $\mathcal{U} = \{u\in\mathbb{R}^3| [-1, -1, -3] \leq u \leq   [1, 1, 9]\}$, $\mathcal{X}_0 = \{x_0\in\mathbb{R}^{6}|[-3, -3, 0.5, 1, 0, 0] \leq x_0 \leq [3, 3, 1, 1, 0, 0]\}$, and $\mathcal{X}_{e0} = \{x_{e0}\in\mathbb{R}^{6}|[-1, -1, -0.4, -1, -1, 0] \leq x_{e0} \leq [1, 1, 1, 1, 1, 0]\}$.

\subsubsection{Two Link Planar Robotic Arm}
The TLPRA is taken from ~\cite{guyananl-2020} and has the state-space representation given by $x=[\theta_1,\theta_2,\dot{\theta}_1,\dot{\theta}_2]^\top$, where the angular position $\theta =[\theta_1,\theta_2]^\top \in \mathbb{R}^2$ and corresponding angular velocities are expressed as $\dot{\theta}=[\dot{\theta}_1,\dot{\theta}_2]^\top \in \mathbb{R}^2$.  The control input $u = [\tau_1,\tau_2]^\top$ corresponds to torque at the two joints. The dynamics are too complex to be approximated by low-degree polynomials hence methods relying on low-degree polynomial approximation of dynamics fail to design a stabilizing controller. The system dynamics are defined as



{\setlength{\arraycolsep}{1pt}
{\begin{equation*}
\small
  \begin{bmatrix}
    \ddot{\theta}_1  \\
    \ddot{\theta}_2   \\
  \end{bmatrix} =
    \begin{bmatrix}
    \alpha + \beta + 2 \eta cos(\theta_2) & \beta + \eta cos(\theta_2)\\
     \beta + \eta cos(\theta_2) & \beta \\
    \end{bmatrix}^ {-1} 
    \begin{bmatrix}
    \tau_1 + \eta(2 \dot{\theta}_1\dot{\theta}_2 + \dot{\theta}_2^2)sin(\theta_2) - \alpha e_1 cos(\theta_1) - \eta e_1 cos(\theta_1 + \theta_2)\\
    \tau_2 - \eta \dot{\theta}_1^2 sin(\theta_2) - \eta e_1 cos(\theta_1 + \theta_2)    \\
    \end{bmatrix} + \begin{bmatrix}
     1  & 0 \\
     0  & 1 \\
    \end{bmatrix}w,
\end{equation*}}}

where $\alpha = (m_1 + m_2)a_1^2$, $\beta = m_2a_2^2$, $\eta = m_2 a_1 a_2$, $e = g/a_1$, $g = 9.8$, $m_1 = 0.8$, $m_2 = 2.3$, $a_1 = 1$ and $a_2 =1$ and $w=[w_1,w_2]^\top\in\mathbb{R}^2$ denotes the disturbance. For the sets, we use $\mathcal{X} = \{x\!\in\!\mathbb{R}^{4}|[-\pi/2,-\pi/2,-\pi/3,-\pi/3] \\\leq x \leq [\pi/2,\pi/2,\pi/3,\pi/3]\}$, $\mathcal{U} = \{u\in\mathbb{R}^2| [0,0] \leq u \leq   [1, 1]\}$, $\mathcal{X}_0 = \{x_0\in\mathbb{R}^{4}|[\pi/2,0,0, 0] \leq x_0 \leq [\pi/2,0, 0,0]\}$, and $\mathcal{X}_{e0} = \{x_{e0}\in\mathbb{R}^{4}|[-0.3, -0.3, 0, 0] \leq x_{e0} \leq [0.3,0.3,0,0]\}$. 

\color{blue}

 \end{document}